\documentclass[12pt]{amsart}
\usepackage{amsbsy,amssymb,amscd,amsfonts,latexsym,amstext,delarray,
amsmath,graphicx} 
\input xypic
\usepackage{url}

\usepackage{color}

\newtheorem{thm}{Theorem}[section]
\newtheorem{prop}[thm]{Proposition}
\newtheorem{cor}[thm]{Corollary}
\newtheorem{lem}[thm]{Lemma}

\numberwithin{equation}{section}

\def\bA{{\mathbb A}}

\def\bG{{\mathbb G}}

\def\bL{{\mathbb L}}

\def\bS{{\mathbb S}}
\def\bT{{\mathbb T}}

\def\C{{\mathbb C}}

\renewcommand{\H}{{\mathbb H}}
\def\N{{\mathbb N}}
\renewcommand{\P}{{\mathbb P}}

\def\R{{\mathbb R}}

\def\cA{{\mathcal A}}
\def\cB{{\mathcal B}}
\def\cC{{\mathcal C}}

\def\cE{{\mathcal E}}

\def\cH{{\mathcal H}}
\def\cI{{\mathcal I}}

\def\cL{{\mathcal L}}

\def\cN{{\mathcal N}}
\def\cO{{\mathcal O}}
\def\cP{{\mathcal P}}

\def\cS{{\mathcal S}}
\def\cT{{\mathcal T}}
\def\cU{{\mathcal U}}
\def\cV{{\mathcal V}}
\def\cW{{\mathcal W}}
\def\cX{{\mathcal X}}

\def\Hom{{\rm Hom}}

\def\cancel#1#2{\ooalign{$\hfil#1\mkern1mu/\hfil$\crcr$#1#2$}}
\def\Dirac{\mathpalette\cancel D}

\title{Gluing Noncommutative Twistor Spaces}
\author{Matilde Marcolli \& Roger Penrose}

\address{California Institute of Technology, Pasadena, CA 91125 \\ USA} 
\address{Mathematical Institute, 
University of Oxford, Oxford
OX2 6GG, United Kingdom}

\begin{document}
\maketitle

\begin{center}
{\it in memory of Sir Michael Atiyah}
\end{center}

\begin{abstract}
We describe a general procedure, based on Gerstenhaber-Schack complexes,
for extending to quantized twistor spaces the Donaldson-Friedman gluing of
twistor spaces via deformation theory of singular spaces. We consider in particular
various possible quantizations of twistor spaces that leave the underlying
spacetime manifold classical, including the geometric quantization of twistor
spaces originally constructed by the second author, as well as some variants based on 
noncommutative geometry.  We discuss specific
aspects of the gluing construction for these different quantization procedures. 
\end{abstract}

\section{Introduction}

\subsection{Introductory historical comments: motivations and signatures}

Twistor theory was originally put forward, in December 1963 (see \cite{Pen67}, \cite{Pen87}), as a novel geometrical proposal for the description of physics, specifically attuned to Einstein's 1905 theory of special relativity. 
Minkowski's 1908 geometrical framework for that theory \cite{Mink} was as a 4-dimensional spacetime $M$ which differs from 
Euclidean $4$-space in that the Euclidean $(+, +, +, +)$-signature metric is replaced by a Lorentzian $(-, +, +, +)$ one, or, 
as we shall prefer here, a Lorentzian $(+, -, - ,-)$ metric, according to which it is the time measure along timelike curves 
that is what is directly defined. The symmetries of Minkowski's spacetime $M$ are given by the $10$-dimensional Poincar\'e group.

\smallskip

In twistor theory, this symmetry is extended to the $15$-dimensional conformal group $SO(2,4)$ of symmetries of 
compactified Minkowski space $M^C$, of topology $S^1\times S^3$, which extends $M$ by the incorporation of 
a ``light cone at infinity" $\cI$ , whose vertex is a point $i$ representing both spatial and temporal infinity, joined 
to a $3$-cylinder of topology $S^2\times \R$ representing ``lightlike" (or null) infinity. 
The free-field Maxwell equations extend to $M^C$, as do the other massless field equations for various spins 
(see \cite{Pen64}, \cite{Pen65}).

\smallskip

In the positive-definite Euclidean case, the connection with physics is less direct, making use of the concept of 
spacetime ``Euclideanization", which plays a role in various approaches to quantum field theory (dating back to \cite{Wick}), 
by means of the ``trick" of allowing the time to be described by an imaginary parameter. The compactification now 
becomes the standard $1$-point conformal compactification of Euclidean $4$-space $\R^4$ to the conformal sphere $S^4$. 
The symmetry group is now $SO(5,1)$. However, passing to the common complexification of both $M^C$ and $S^4$, 
we again obtain a complex-conformal $4$-quadric $\C S^4$, wherein the single point $i$, in the compactification of $S^4$ 
now extends to the complex $3$-cone $\C\cI$, with vertex $i$.

\smallskip

The idea of twistor theory is to appeal to the Grassmann--Klein representation of the family of projective lines in 
complex projective $3$-space $\C\P^3$ as a complex $4$-quadric, but where we now take this in the reverse sense; 
that is to say, the complexified conformal spacetime $\C S^4$ is to be regarded as the Klein representation of 
projective lines in a complex projective $3$-space $\P\cT$, the projective space of the complex vector space $\cT$, 
referred to as twistor space. The complexifications of $S^4$ and $M^C$ are identical, so in each case we get the 
same projective twistor space $\P\cT$. However we are also interested in reality structures in the two cases, and
these come out very differently, despite the fact that the same complex space $\C\P^3$ arises as its twistor space 
$\P\cT$ in each case. The difference lies in the way that the ``real" points of the spacetime are interpreted 
within $\P\cT$.

\smallskip

Let us first consider the original Lorentzian case \cite{Pen67}.  Here we obtain a realization of the isomorphism 
between the Minkowskian conformal spacetime group $SO^+(4,2)$ and the twistor symmetry group $SU(2,2)$. 
The points of the compactified spacetime $M^C$ are all represented by $\C\P^2$s lying in a $5$-real-dimensional 
subspace $\P\cN\subset \P\cT$ of topology $S^3\times S^2$. The points of $\P\cN$ themselves correspond to 
null straight lines--i.e.~light rays--in $M^C$, including the generators of $\cI$. The $S^2$-family of light rays 
through a fixed point $p$ in $M^C$ corresponds to a $\C\P^1$ in $\P\cN$. The condition for two points of $M^C$ 
to be null separated (i.e.~joined by a light ray) is that the $\C\P^1$s that represent them in $\P\cN$ intersect.

\smallskip

The points of $\P\cT \smallsetminus \P\cN$ also have an interpretation within the (compactifed) real spacetime $M^C$, 
but not as light rays. It is easiest to interpret a point $p$ in $\P\cT \smallsetminus \P\cN$ by considering, instead, the dual $p^*$, 
with respect to the $SU(2,2)$ structure, which is a $\C\P^2$ that intersects $\P\cN$ in an $S^3$. This $S^3$ can be taken 
to represent $p$ via this duality. It is naturally fibred according to the Clifford--Hopf fibration $S^1 \hookrightarrow S^3 \to S^2$,
where each point $q\in S^3$ lies on an $S^1$ fibre which is the intersection of $S^3$ with the complex line joining $q$ to $p$. 
In spacetime terms, this construction provides us with the realization of $S^3$ as a twisting $3$-parameter family of light 
rays termed a ``Robinson congruence", which originally provided the name ``twistor" (see \cite{Pen87}).

\smallskip

In the case of a Riemannian (Euclideanized) spacetime $S^4$, where we are primarily interested only in the 
conformal-Riemannian structure of $S^4$, each point of $S^4$ still corresponds to a $\C\P^1$ in the $\C\P^3$ (that is $\P\cT$), 
but in this Riemannian case we have no null-separated points in the spacetime. Accordingly, none of the lines in this
$\C\P^3$ (i.e.~$\P\cT$) can intersect, and thus instead of the real spacetime points being determined by a subspace 
(i.e.~$\P\cN$) within $\P\cT$, we have a fibration
$$ \C\P^1 \hookrightarrow \C\P^3 \to S^4 $$
(see, for example \cite{ADM}).

\smallskip

The twistor framework for either spacetime signature, or in the complex case, becomes remarkably useful for the 
description of massless free fields (in regions within $S^4$, $M^C$, or their common complexification $\C S^4$) 
for various spins, such as the free electromagnetic Maxwell field, which is the case of spin $1$. Any such field--which 
we may now take to be a complex solution of the field equations (this being relevant to quantum wavefunctions)--can 
be separated into its right-handed (positive helicity $s$) and left-handed (negative helicity $s$) parts, where $|s|$ 
is the spin of the field. The quantity $s$, where $2s$ is necessarily an integer, $s$ being called the helicity, is
negative for left-handed helicity and positive for right-handed, and we also allow $s=0$. The field equation for $s=0$ 
is simply the wave equation and for $|s|=1$, we get the Maxwell equations. For $|s|=2$, we get the description of the 
free gravitational field
according to the weak-field linear limit of Einstein's general theory of relativity. For $s=\pm 1/2$ we get the 
neutrino/antineutrino equations in the limiting case of zero mass.

\smallskip

Explicitly, the solutions of these equations are represented very directly as simple contour integrals of holomorphic 
functions of a single twistor ${\bf z}$, referred to as twistor functions, taken to be holomorphic and homogeneous 
of degree $-2s -2$, but otherwise subject to no equations. Twistor finctions are more properly thought of as defining 
elements of $1$st \v{C}ech cohomology (see, for example, \cite{PenRind}, particularly \S 6.10). To describe the field 
in a local region of the spacetime, it is sufficient to use a $2$-set \v{C}ech covering of the corresponding region in 
$\P\cT$, which we may regard as an open ``thickening" of a $\C\P^1$.

\smallskip

In the case $s = -2$, so that the twistor function's homogeneity is $+2$, we find that this construction generates the left-handed 
(i.e.~negative helicity) weak-field solutions of the full Einstein vacuum field equations (with or without a cosmological 
constant $\Lambda$), i.e.~(complex) Einstein $4$-manifolds. This comes about as follows. In the linear case, locally (in the spacetime) 
we need consider only a $2$-patch \v{C}ech covering, the twistor function being defined on their overlap. For the full non-linear 
situation, we take this twistor function to define a ``gluing" of one patch to the other which differs from the identity map. 
A theorem of Kodaira \cite{Kod} and Kodaira--Spencer \cite{KodSp} tells us that, so long as this displacement is not too large, we still get a $4$-parameter family of $\C\P^1$s, straddling the patches. Regarding these as defining points in a conformal $4$-manifold (with conformal structure defined in terms of intersections between $\C\P^1$s, as in the conformally flat case), we obtain, locally, completely general analytic conformal $4$-manifolds which are anti-self-dual, this constraint referring to the vanishing of the self-dual part ${\bf W}^+$ of the Weyl curvature tensor ${\bf W}={\bf W}^+ + {\bf W}^-$, where ${\bf W}^-$ would be its anti-self-dual part. This construction yields, locally, the most general such anti-self-dual conformal $4$-manifold.

\smallskip

Moreover, in standard twistor space $\cT$, we have a certain $2$-form ${\bf I}$ referred to as he ``infinity twistor", which gives $M$ a Euclidean metric if ${\bf I}$ is degenerate (i.e.~rank $2$) and a de Sitter or anti-de Sitter one if ${\bf I}$ is non-degenerate. If in the above ``gluing" we preserve ${\bf I}$ from patch to patch, then the assigned metric is necessarily Einstein (Ricci tensor being proportional to the metric tensor) and provides us, locally, with the most general such anti-self-dual $4$-space (see \cite{Pen67}, \cite{Ward}). 
This has become known as the ``non-linear graviton" construction.

\smallskip

The question naturally arises as to whether some sort of twistor construction might give rise to generic space-times 
which provide us with completely general (analytic) space-times. Even more pertinently, in the directly physical 
Lorentzian case, one might well regard the above construction as completely useless because in this case the 
decomposition ${\bf W}={\bf W}^+ + {\bf W}^-$ is a complex one, where ${\bf W}^+$ and ${\bf W}^-$ are complex 
conjugates of one another, so that if one vanishes, so does the other, and we are restricted to conformally flat space-times.

\smallskip

On the other hand, one might imagine that there could be some ``non-linearization" of the twistor procedure that had 
enabled us to generate linearized self-dual Weyl tensors from twistor functions of homogeneity degree $-6$, analogously 
to the way that the non-linear graviton achieved this for homogeneity $+2$. (This had been termed the ``googly problem", 
by analogy with a difficult bowling action in the game of cricket.) Then perhaps one might ``add the two together" in some 
sense so as to obtain a solution to the general problem of finding a full expression for solutions of Einstein's vacuum 
equations in twistor terms. However, despite many attempts, employing different types of idea, no solution has yet 
come close to a solution along these lines.

\smallskip

It should be mentioned, at this point that many researchers have adopted a different viewpoint, referred to as the 
use of ``ambitwistors", which involves combining a twistor with a dual twistor into a single entity. Although much 
significant work has been achieved along these lines (see, for example \cite{LeBrun}), nevertheless, this must 
be regarded as a solution to a different kind of problem, and some of the economy that is a feature of twistor theory is lost.

\smallskip

Another way of thinking about our difficulty here is that spatial reflection takes twistors into dual twistors (members 
of the dual twistor space); moreover self-dual and anti-self-dual parts of the Weyl curvatures are interchanged 
upon $3$-spatial reflection. In the Lorentz-signature framework, the complex conjugate of a twistor is a dual twistor, 
and vice versa. Accordingly, a holomorphic function of a twistor would reflect into an anti-holomorphic one, so it would 
seem that we need to extend the formalism to include both twistors and dual twistors if we are to preserve holomorphicity, 
so we appear to be driven back to ambitwistors, and thereby lose much of the economy that is inherent in the twistor formalism.

\smallskip

However, there is another solution, which is to appeal to the framework of {\em twistor quantization}, whereby holomorphicity (crucial for many expressions of twistor theory) is retained, although at the expense of the non-commutativity that is inherent in the use of differential operators. In this procedure, the complex conjugate of a twistor is replaced by a holomorphic differential operator \cite{Pen68}, this procedure having had relevance in many of the expressions of twistor theory. Some aspects of this non-commutative algebraic approach, in the Lorentzian framework, are to appear elsewhere, under the name of ``palatial" twistor theory (see \cite{Penrose2020}), but here we go more deeply into the general structure of this procedure, mainly in the positive-definite signature situation, and explore the resulting non-commutative twistor geometry describing classical spacetimes.

\subsection{Summary and organization of the paper}

In \cite{Pen68} one of us introduced a quantization of twistor spaces, based on
a geometric quantization procedure. This construction was motivated by the
fact that extending the twistor formalism for conformally curved spacetimes
involves the problem of dealing with non-analytic transformations of twistor
space that mix the $Z^\alpha$ and the conjugate $\bar Z_\alpha$ coordinates. 
The observation that such transformation do, however, preserve Poisson
brackets obtained by viewing the $\bar Z_\alpha$ as canonical conjugate 
variables of the $Z^\alpha$ leads naturally to considering a quantized 
version of twistor space, which still makes it possible to work with
holomorphic functions of the $Z^\alpha$, where the operator corresponding
to $\bar Z_\alpha$ is identified with $\partial/\partial Z^\alpha$. 

\smallskip

More recently, noncommutative deformations of twistor spaces
were considered in the context of noncommutative geometry
(see \cite{BrLa}, \cite{BrMa}, \cite{LaPaRe}, \cite{LaWvS}), obtained using the
Connes--Landi $\theta$-deformation technique \cite{CoLa}.
These constructions are based on quantizing the Hopf fibration
\begin{equation}\label{HopfCP3S4}
 \C\P^1 \hookrightarrow \C\P^3 \rightarrow S^4. 
\end{equation}
However, all these constructions involve a 
quantization of the spacetime manifold $S^4$ and a compatible
quantization of the twistor space $\C\P^3$ determined by the
geometry of the Hopf fibration. The motivation for these noncommutative
deformations lies primarily in the construction of instantons on noncommutative
$4$-spheres, hence the noncommutative deformation of the spacetime
manifold is crucial to the purpose. 

\smallskip

The point of view we are interested in here is different, in the
sense that we are interested in quantizations of the twistor space that
leave the spacetime manifold commutative. 

\smallskip

The main focus of the paper is the gluing problem for
quantized twistor spaces formulated in Part D of \cite{Penrose2020}.
We use the Gerstenhaber--Shack theory of noncommutative deformations \cite{GeSch}
to extend the Donaldson--Friedman gluing \cite{DoFr} of classical twistor spaces
to their noncommutative counterparts. In this paper we work primarily with Riemannian,
rather than Lorentzian manifolds, in order to be able to directly compare the gluing result 
we discuss with the classical result of \cite{DoFr}. However, the general procedure 
we describe for the gluing of quantized twistor spaces would apply also in the
Lorentzian setting in which the problem was originally formulated in \cite{Penrose2020}.

\smallskip

Section~\ref{NCtwistSec} of the paper introduces some examples of noncommutative
deformations of twistor spaces.  In particular, we show that, in addition to the
geometric quantization construction of quantized twistor spaces
originally introduced by one of us in \cite{Pen68}, other variants are possible, which
have a natural interpretation in the setting of noncommutative geometry.
In particular, we investigate quantizations of twistor spaces that are obtained,
for an (anti)self-dual Riemannian manifold $M$, by imposing that $M$ remains
commutative and that the quantization of the twistor space $Z(M)$ is
compatible with the Hopf fibrations relating the twistor space $Z(M)$ to $M$
with twistor lines $\C\P^1$ fibers and the twistor space $Z(M)$ and the sphere
bundle $S(M)$ of the spinor bundle $\cS^+(M)$ with fiber $S^1$.
We describe the geometry of some possible quantizations obtained via 
these requirements. We also show the different role that the Hopf fibration
plays in the geometric quantization of twistor spaces of \cite{Pen68}, in the Lorentzian setting,
and its compatibility with the quantization. 

\smallskip

In Section~\ref{DefGlueSec} we focus on the main question of gluing of
noncommutative twistor spaces. We present an abstract and very general procedure
that applies to any chosen noncommutative deformation that can be described in terms
of deformation quantization. In Section~\ref{GlueCasesSec} we show more explicitly
how the examples of twistor space quantization introduced in Section~\ref{NCtwistSec}
fit into this general procedure.

\smallskip

Our main construction of Section~\ref{DefGlueSec} is based on a
noncommutative generalization of the gluing result of \cite{DoFr}.
Donaldson and Friedman showed in \cite{DoFr} that one can associate
to the connected sum $M= M_1 \# M_2$ of two (anti)self-dual Riemannian 
$4$-manifolds $M_i$ a singular space $\tilde Z(M)$ obtained by
first blowing up the twistor spaces $Z(M_i)$ along one of the
$\C\P^1$ fibers and then gluing together the two exceptional 
divisors, $\tilde Z(M)=\tilde Z(M_1) \sqcup_{E_1\simeq E_2} \tilde Z(M_2)$,
with $\tilde Z(M_i)={\rm Bl}_{\C\P^1}(Z(M_i))$. The gluing map of the
exceptional divisors is determined by an orientation-reversing isometry
of the tangent spaces of $M_i$ at the points $x_i$ where the connected
sum is performed and where the respective fibers $F_{x_i}=\C\P^1$ 
are blown up. The space $\tilde Z(M)$
obtained in this way has a normal crossing singularity along the identified
exceptional divisors, which form a $\C\P^1 \times \C\P^1$.  In \cite{DoFr} they then
consider the question of whether the singular space $\tilde Z(M)$ admits
an unobstructed deformation to a smooth space, and they show that,
when this is the case, the resulting smooth space is the twistor
space $Z(M)$ of the connected sum manifold. In particular, this ensures
the existence of (anti)self-dual metrics on the connected sum.  
The analysis of deformations and obstructions used for the result of \cite{DoFr} 
is based on a deformation theory of spaces with normal crossings singularites
developed in \cite{Fri}. The main deformation result of \cite{DoFr}  states that, 
if the twistor spaces $Z_i=Z(M_i)$ have unobstructed deformation theory,
namely if the cohomology $H^2(Z_i, \cO(TZ_i))=0$, then the deformation theory
of $\tilde Z(M)$ is also unobstructed. 

\smallskip

We investigate to what extent the gluing and deformation procedure of
\cite{DoFr} can be adapted to quantized twistor spaces. 
The work of Gerstenhaber and Schack \cite{GeSch} showed that 
classical Kodaira--Spencer deformation theory of complex manifolds
can be subsumed as a ``commutative part" of a more general
deformation theory that includes noncommutative deformations and
that is governed by a parameterization of infinitesimal deformations
and obstructions in terms of Hochschild cohomology. 

\smallskip

Using this formulation of deformation theory, and
starting with unobstructed noncommutative deformations of the twistor 
spaces $Z_i=Z(M_i)$, relative to a choice of a twistor line $L_i$, we
show that it is possible to obtain an unobstructed noncommutative
deformation of the singular space $\tilde Z$, subject to a compatibility
condition between the choices of the cochains that define the higher
terms of the deformation. In particular, if the commutative parts of
the deformations of the $Z_i$ is unobstructed, the construction recovers
the gluing and deformation of \cite{DoFr}, so that we can identify the
resulting noncommutative deformation of $\tilde Z$ with a noncommutative
deformation of the twistor space $Z(M_1\# M_2)$ when the latter exists.
If the commutative part of the deformations  of the $Z_i$ is obstructed
but the noncommutative deformations are unobstructed, the resulting
noncommutative deformation of $\tilde Z$ can be viewed as a quantized
twistor space for $M_1\# M_2$ which may exist even if the classical
one does not, for instance in cases when $M_1\# M_2$ does not carry 
an (anti)self-dual structure. 

\smallskip

In Section~\ref{GlueCasesSec} we look again at the specific
examples of noncommutative deformations of twistor spaces
discussed in Section~\ref{NCtwistSec} and we show to
what extent the general gluing procedure of Section~\ref{DefGlueSec}
applies in each case. We show that it can be applied to the
original quantization of twistor spaces of \cite{Pen68}, where it agrees
with a geometric quantization of a Gompf sum of symplectic 
manifolds. We then show how the gluing works explicitly for
the other variants of quantization of twistor spaces obtained in Section~\ref{NCtwistSec}
from deformations of the Hopf fibration, with different geometric properties of the corresponding
deformation theory.

\section{Noncommutative Twistor Spaces}\label{NCtwistSec}

We discuss different noncommutative deformations of twistor spaces.
Our primary interest is the quantized twistor space introduced by one of us in \cite{Pen68}.
However, we also show that, if one works with (anti)self-dual Riemannian manifolds
and imposes the requirements that the noncommutative deformation of the twistor
space is compatible with the Hopf fibration, while leaving the spacetime manifold
commutative, this can lead to a choice of somewhat different quantizations of the twistor spaces.
In particular we first recall the geometric quantization of twistor space,
viewed in the context of geometric quantization and deformation quantization,
and then we analyze a few different variants of the construction of a quantized
twistor space.  We then return to discuss the geometric quantization
of twistor spaces of \cite{Pen68} and we analyze more in detail the role of the Hopf fibration and 
the compatibility of the quantization with the Hopf fibration, which is different from
the other cases we discuss in this section. 

\smallskip

In order to construct these different quantizations of twistor space, we 
focus on the geometry of the Hopf fibration. 
We show that there are different ways of deforming the Hopf fibration
$S^1\hookrightarrow S^3\to S^2$ to noncommutative spaces,
which result in a compatible noncommutative deformation of
the Hopf fibration $S^3\hookrightarrow S^7 \to S^4$, and more
generally of the unit sphere bundle $\bS(\Lambda_+(M))$ of
a self-dual $4$-manifold $M$, in a way that leaves the space manifold
$S^4$ or $M$ commutative. 

\smallskip

A first method we discuss is based on deforming the Hopf fibration
$S^1\hookrightarrow S^3\to S^2$ by deforming all the $2$-tori
of the Hopf foliation of $S^3$ to noncommutative tori. This
deformation and the resulting deformations of $\bS(\Lambda_+(M))$
fall within the setting of the Connes--Landi $\theta$-deformations
of noncommutative geometry. Moreover, they have a counterpart,
where the noncommutative deformation can be expressed in
terms of a noncommutative deformation of the fibrations
$\C^* \hookrightarrow \C^2\smallsetminus \{ 0\} \to \C\P^1$ 
and $\C^* \hookrightarrow \C^4\smallsetminus \{ 0\} \to \C\P^3$,
and which can be described in terms of the noncommutative
toric deformations of Cirio--Landi--Szabo, \cite{CiLaSza1}, \cite{CiLaSza2}, \cite{CiLaSza3}.
Hovewer, we will show that this method does not correspond to the quantization
of twistor spaces introduced in \cite{Pen68}. Indeed, in this deformation the
base $S^2\simeq \C\P^1$ of the Hopf fibration $S^1\hookrightarrow S^3\to S^2$
remains commutative, unlike what is expected as effect of the quantization of \cite{Pen68}.
This results in a noncommutative sphere bundle $\bS(\Lambda_+(M))_\theta$ that
fibers with $S^1$-fibers over a commutative twistor space $Z(M)$ and also fibers
over the commutative spacetime manifold $M$,  with fibers the noncommutative 
spheres $S^3_\theta$ .

\smallskip

A second method is based instead on a noncommutative
deformation of the Hopf fibration
$S^1\hookrightarrow S^3\to S^2$ that is based on the
deformation quantization method originally introduced in \cite{BFFLS}, where
the compatibility of the deformation and the Hopf fibration is achieved using
the construction of  \cite{OMYM1}, \cite{OMYM2}. 
We will show that this noncommutative deformation
induces a deformation of the sphere bundle 
$S^3 \hookrightarrow \bS(\Lambda_+(M)) \to M$ that leaves the self-dual $4$-manifold
$M$ commutative, and a compatible noncommutative deformation of the twistor space 
$\C\P^1\hookrightarrow Z(M) \to M$. The resulting noncommutative $Z(M)_\hbar$
obtained in this way, however, is not exactly the quantization described in \cite{Pen68}. 
Indeed we show that, instead of the commutation relations $[Z^\alpha, Z^\beta]=0$,
$[\bar Z_\alpha, \bar Z_\beta]=0$ and $[Z^\alpha, \bar Z_\beta]=\hbar \delta^\alpha_\beta$,
in the twistor space $Z(M)_\hbar$ we obtain by this deformation method, both $[Z^\alpha,\bar Z_\alpha]=\hbar$
and also $[Z^\alpha,\overline{Z^\alpha} ]=\hbar$, where $\bar Z_0=\overline{Z^2}$, $\bar Z_1 =\overline{Z^3}$,
$\bar Z_2 =\overline{Z^0}$ and $\bar Z_3 =\overline{Z^1}$. 

\smallskip

This variant of the commutation relations of \cite{Pen68}, with the additional non-trivial
commutators $[Z^\alpha,\overline{Z^\alpha} ]=\hbar$, also has a natural interpretation
in terms of the settings described by one of us in \cite{Penrose2020}. Indeed, in this case
one is considering in the deformation both the symplectic form of \cite{Pen68} (see \eqref{symplectic} below),
as well as the one discussed in Section C.6 of \cite{Penrose2020} and related to the
cosmological constant. 

\smallskip

There is also a third construction that we will discuss, which is also associated to 
deformation quantization methods and which produces a noncommutative twistor space that is
an almost-commutative geometry (in the sense of \cite{Cac2}) over the
spacetime manifold $M$. This construction has as base of the noncommutative
Hopf fibration the fuzzy $2$-sphere. We will also discuss briefly the properties of
these resulting ``fuzzy twistor spaces". In this case also the commutation relations between
the twistor variables are as in the previous case, rather than as in \cite{Pen68}.

\smallskip

By focusing on the case of $M=S^4$ we then show that, if we require the same form of compatibility
with the Hopf fibration as in the previous cases, then the commutator prescription 
\begin{equation}\label{relQtwistor}
[Z^\alpha, Z^\beta]=0, \ \ \  
[\bar Z_\alpha, \bar Z_\beta]=0, \ \ \  [Z^\alpha, \bar Z_\beta]=\hbar \delta^\alpha_\beta. 
\end{equation}
of \cite{Pen68} would imply that the spacetime manifold $S^4$ is also deformed
to a noncommutative space. 
In the original construction of \cite{Pen68}, however, the role of the Hopf
fibration is different from the other cases we discuss in this section, and is
best understood in the original Lorentzian setting. We describe in Section~\ref{PenroseQHopfSec} 
how copies of the Hopf fibration $S^1\hookrightarrow S^3 \to S^2$ are embedded in
the subspace $\P N$ of the twistor space defined by the vanishing of the signature $(+,+,-,-)$
norm $\sum_\alpha Z^\alpha \bar Z_\alpha$ of the $SU(2,2)$ structure on $\C\P^3$. We then
show that the geometric quantization of twistor space induces a compatible quantization of
these Hopf fibrations.  This different role of the Hopf fibration then suggests yet another
possible variant of noncommutative deformation of twistor space, again based on the 
$\theta$-deformations, applied to all the Hopf fibrations in $\P N$. We discuss this
other variant in Section~\ref{SecondThetaSec}.

\smallskip

While our primary interest is in investigating the gluing problem for the original
geometric quantization of twistor space of \cite{Pen68}, we include the discussion
of all these different noncommutative deformation methods anyway,
because it seems interesting to compare how these constructions behave
with respect to the gluing problem, see Section~\ref{GlueCasesSec}.

\subsection{Symplectic geometric quantization of twistor space}\label{PenroseQSec}

We review briefly the quantization of twistor spaces originally introduced by one of us in \cite{Pen68}, seen
in terms of symplectic geometric quantization and in terms of deformation
quantization. The deformation quantization viewpoint will be useful in order
to relate this noncommutative deformation to the general deformation and
obstruction procedure for the gluing of noncommutative twistor spaces
that we introduce in Section~\ref{DefGlueSec}.

\smallskip

Let $(M,g)$ be a Riemannian manifold with an (anti)self-dual metric. Then there is an
integrable almost complex structure $J$ on the tangent bundle $TZ$ of the 
twistor space $Z=Z(M)=\bS(\Lambda_+(M))=\P(\cS^+(M))$, with $\Lambda_+(M)$
the bundle of self-dual $2$-forms and $\cS(M)$ the positive part of the spinor bundle,
and $Z$ is a $3$-dimensional complex manifold. The fibration 
$\C^*\hookrightarrow \cS^+(M)_0 \to \P(\cS^+(M))$, with $\cS^+(M)_0$ the complement 
of the zero section, in turn determines a complex involution $J$ on $\cS^+(M)_0$.
We denote by $Z^\alpha$, $\alpha=0,\cdots,3$ and $\bar Z_\alpha$ the
complex coordinates that are conjugate under this complex structure. We can
consider the symplectic form
\begin{equation}\label{symplectic}
\omega = \sum_\alpha dZ^\alpha \wedge d \bar Z_\alpha. 
\end{equation}
The complex structure $J$ determines subspaces $T^{0,1}$ and $T^{1.0}$ of
the complexified $T(\cS^+(M)_0)^\C$, spanned by vectors $v\pm i J v$. The subspace
$P=T^{0,1}$, spanned by the vectors $\partial/\partial \bar Z_\alpha$ gives the complex 
polarization used for geometric quantization. 

\smallskip

The geometric quantization procedure, associated to a symplectic manifold $(X,\omega)$,
consists of two steps: the prequantization and the polarization and quantization, \cite{Wood}.
In the prequantization stage one considers the Hilbert space of square-integrable
sections of a hermitian line bundle $\cH=L^2(X,\cL)$, with Chern class $c_1(\cL)=\hbar^{-1}[\omega]$,
with a prequantization map that assigns to functions $f$ on $X$ operators on $\cH$
of the form $-i\hbar X_f - \theta(X_f) +f$, with $\theta$ the symplectic potential and $X_f$
the Hamiltonian vector field associated to $f$, so that Poisson brackets of functions are mapped, up to a factor of $i\hbar^{-1}$ to commutators of operators. Here $d - i \hbar^{-1} \theta$ is the local form
of a connection on the line bundle $\cL$. For $X= T^*\R^N$ with coordinates $(q^k,p_k)$
the operators assigned to the position coordinates $q^k$ are of the form $i\hbar \frac{\partial}{\partial p_k} + q^k$
and those associated to the momenta $p_\ell$ are of the form $-i \hbar \frac{\partial}{\partial q^\ell}$. 
The prequantization space and operators involve functions of a mixture of both positions and momenta. 
This can be narrowed down, through a choice of polarization, 
to a set of variables that separates positions and momenta and makes it possible to
work with just half of the variable. The polarization $P$ is a half-dimensional subbundle
of $TX$. The prequantum sections of $\cL$ are then replaced by the polarized sections, namely
those that are, in the appropriate sense, covariantly constant along $P$. 

\smallskip

More explicitly, in the case 
we are considering, we write the symplectic form \eqref{symplectic} as
$\omega =i \partial \bar\partial K$ with $K=\sum_\alpha Z^\alpha \bar Z_\alpha$
and we consider the symplectic potential $\theta = -i \partial K$ that vanishes on
the polarization $P$. Thus polarized sections of the prequantum line bundle can
be identified in a local chart with functions of the holomorphic coordinates $Z^\alpha$.
The operators corresponding in this quantization to the coordinates $Z^\alpha$ and
$\bar Z_\alpha$ satisfy the commutator relations \eqref{relQtwistor}.

\smallskip

For our purpose of investigating the gluing of quantized twistor spaces, it is convenient
to associate to this description of the quantized twistor space in terms of symplectic
quantization a description in terms of deformation quantization. This can be done
along the lines of Fedosov quantization, \cite{Fed}.

\smallskip

In the theory of deformation quantization developed in \cite{BFFLS}, 
a formal deformation of $\cA$ is a $\C[[t]]$-algebra
obtained by assigning a $\C[[t]]$-linear multiplication $\alpha_t: \cA[[t]]\times \cA[[t]] \to \cA[[t]]$, with
$\alpha_t=\alpha + t\alpha_1 + t^2 \alpha_2 + \cdots$ with $\alpha$ the multiplication of $\cA$ and
$\C$-linear maps $\alpha_i : \cA\times \cA \to \cA$, so that associativity 
$\alpha_t(\alpha_t(a,b),c)=\alpha_t(a,\alpha_t(b,c))$ holds. 

\smallskip

Under the procedure of Fedosov quantization \cite{Fed}, given a symplectic manifold $(X,\omega)$,
one considers the space $\cW={\rm Sym}^*_\C(TX)[[\hbar]]$, with the two gradings of symmetric powers 
and of powers of $\hbar$, and the subspace of flat sections $\Gamma_\nabla(\cW)$
with respect to a flat connection $\nabla$ on $\cW$, whose potential can be recursively determined as
power series with respect to both gradings (see \cite{Fed}, \cite{Fed2}, \cite{Nolle}). Fedosov showed
that every $f\in \cC^\infty(X)[[\hbar]]$ determines uniquely a section $\rho(f)=\hat f$ in $\Gamma_\nabla(\cW)$
with degree zero part (in the ${\rm Sym}^*_\C(TX)$-grading) equal to $f$ and that the associative
product of the deformation quantization $\cC^\infty(X)[[\hbar]]$ can be obtained by inverting
this map, $f\star_\hbar g =\rho^{-1}(\hat f \circ \hat g)$. 

\smallskip

\begin{prop}\label{geomdeftwistor}
The geometric geometric quantization of twistor spaces of \cite{Pen68} has a compatible associated
deformation quantization. 
\end{prop}

\proof
In the case of geometric quantization on a complex manifold, with the holomorphic polarization,
as in our case of twistor spaces, it is shown in \cite{Nolle} that the compatibility between 
geometric quantization and deformation quantization reduces to two conditions:
\begin{enumerate}
\item For $f,g$ holomorphic, the product $f \star_\hbar g$ is also holomorphic.
\item The $\star_\hbar$ product of a function that is affine-linear in $\frac{\partial K}{\partial Z^\alpha}$
with a holomorphic function is still affine-linear.
\end{enumerate}
Thus, it suffices to show that these two conditions are satisfied to ensure that the geometric
quantization of twistor spaces can also be described in terms of an associated compatible
deformation quantization. We have $K=\sum_\alpha Z^\alpha \bar Z_\alpha$, hence the
second condition requires that the product of an affine-linear function of the $\bar Z_\alpha$
coordinates with a holomorphic function of the $Z^\alpha$ coordinates is still 
affine-linear in the $\bar Z_\alpha$. The coefficients $\frac{\partial^2 K}{\partial Z^\alpha \partial \bar Z_\beta}=\delta_{\alpha,\beta}$ of the symplectic form are constant, hence there are no associated
curvature terms. In this case, as observed in \cite{Fed}, the product takes the form
$$ f \star_\hbar g =\exp (-\frac{i \hbar}{2} \omega^{\alpha\beta} \frac{\partial}{\partial X^\alpha} \frac{\partial}{\partial Y^\beta} ) f(X,\hbar) g (Y,\hbar) |_{X=Y} $$
$$ = \sum_{k=0}^\infty (\frac{-i \hbar}{2})^k \frac{1}{k!} \omega^{\alpha_1\beta_1}\cdots \omega^{\alpha_k\beta_k}
\frac{\partial^k f}{\partial X^{\alpha_1}\cdots \partial X^{\alpha_k}} \frac{\partial^k g}{\partial X^{\beta_1}\cdots \partial X^{\beta_k}}. $$
It is then clear that, if both $f$ and $g$ are holomorphic functions of the holomorphic
coordinates $Z^\alpha$ then also $f\star_\hbar g$ is a holomorphic function of the $Z^\alpha$
and if $f$ is affine linear in the $\bar Z_\alpha$ and $g$ is a holomorphic function of the
$Z^\alpha$, the product still has an affine-linear dependence on the $\bar Z_\alpha$ variables. 
\endproof

\smallskip
\subsection{Hopf fibrations and twistor spaces}\label{twistHopfSec}

We discuss next some other possible quantizations of twistor spaces,
which also have the property that the underlying spacetime manifold
remains classical. These are obtained using different methods
in noncommutative geometry (given, respectively, by $\theta$-deformations,
deformation quantization, and fuzzy spaces), applied to the Hopf
fibration $S^1\hookrightarrow S^3 \to S^2$. In order to describe
these quantizations, we first recall a few facts regarding the role of the 
Hopf fibrations in the geometry of twistor spaces.

\smallskip

The first significant example of twistor space that illustrates the relation to
the Hopf fibration is the case of $M=S^4$ with $Z(M)=\C\P^3$ and the Hopf fibration 
\eqref{HopfCP3S4} relating them. 
The twistor space construction is illustrated in this case by the 
commutative diagram of Hopf fibrations
\begin{equation}\label{diagHopfCP3}
\xymatrix{ S^1 \rto^{=} \dto & S^1 \dto &  \\ 
S^3 \rto \dto & S^7 \rto \dto & S^4 \dto^{=} \\ \C\P^1 \rto & \C\P^3 \rto & S^4 =\H\P^1 }
\end{equation}

\smallskip

The Hopf fibration projection $S^3\to S^2$ is given by $(z_0,z_1)\mapsto (2 z_0 \bar z_1, |z_0|^2 - |z_1|^2)$
or, in Hopf coordinates, by $(e^{i\xi_1}\cos\eta, e^{i\xi_2}\sin\eta)\mapsto (e^{i(\xi_1-\xi_2)} \sin 2\eta, \cos 2\eta)$.
In fact, after the identification of $S^2$ with $\C\P^1$ via the stereographic projection, the Hopf projection 
map is simply the restriction to $S^3\subset \C^2\smallsetminus \{ 0 \}$ of the projection 
$\C^2\smallsetminus \{ 0 \} \to \C\P^1$,
$(z_0,z_1)\mapsto (z_0:z_1)$, in the affine chart 
$(z_0,z_1)\mapsto z_0 z_1^{-1}$. In this form, the Hopf projection map remains of the same form 
$(q_0,q_1)\mapsto q_0 q_1^{-1}$ in the case of the Hopf fibration $S^3\hookrightarrow S^7 \to 
S^4\simeq \H\P^1$, after replacing $z_i\in \C$ by quaternions $q_i\in \H$. Thus, we can
equivalently consider the diagram of fibrations
\begin{equation}\label{diagCHopfCP3}
\xymatrix{ \C^* \rto^{=} \dto & \C^* \dto &  \\ 
\C^2 \smallsetminus \{ 0 \} \rto \dto & \C^4 \smallsetminus \{ 0 \} \rto \dto & \H\P^1 \dto^{=} \\ \C\P^1 \rto & \C\P^3 \rto & \H\P^1 }
\end{equation}
where we identify $\C^2 \smallsetminus \{ 0 \}=\H\smallsetminus \{ 0 \}$ and $\C^4 \smallsetminus \{ 0 \}=\H \times \H 
\smallsetminus \{ 0 \}$.

\smallskip

Our investigation of different forms of quantization of twistor spaces starts by
considering the  twistor space $\C\P^3$ and possible quantizations of 
the Hopf fibration $S^1 \hookrightarrow S^3 \to S^2$ that leave the spacetime
$S^4$ manifold classical. 
This will have, in particular, the advantage that the same
method can be applied to more general spacetime manifolds $M$, in both the Lorentzian
and Euclidean setting, that admit a twistor space $Z=Z(M)$ with a corresponding
fibration
\begin{equation}\label{ZMfib}
\C\P^1 \hookrightarrow Z(M) \rightarrow M . 
\end{equation}
In this more general setting, we want to consider a diagram analogous to the
diagram \eqref{diagHopfCP3} relating the Hopf fibration of the
twistor space $\C\P^3$ to the Hopf fibrations $S^1 \hookrightarrow S^3 \to S^2$ and
$S^3 \hookrightarrow S^7 \to S^4$.

\smallskip

More precisely, let $M$ be an (anti)-self-dual Riemannian $4$-manifold. 
Then the associated twistor space $Z=Z(M)$ is the sphere bundle $Z(M)=\bS(\Lambda_+(M))$ 
of $\Lambda_+(M)$, where $\Lambda^2(M)=\Lambda_+(M)\oplus \Lambda_-(M)$ is
the splitting of $2$-forms into self-dual and anti-self-dual parts. Thus,  
there is a fibration $S^2\hookrightarrow Z\to M$, see \cite{AHS}. If $\cS(M)=\cS^+(M)\oplus \cS^-(M)$
denotes the spinor bundle of $M$, with $\cS^\pm(M)$ complex $2$-plane bundles, then one
can also describe the twistor space as the projectivized spinor bundle $Z(M)=\P(\cS^+(M))$.
The self-duality condition guarantees integrability of the almost complex structure \cite{AHS}, hence
the twistor space is a complex manifold (in general non-K\"ahler, unless $M$ is conformally equivalent
to either $S^4$ or $\C\P^2$, \cite{Hitchin}); the  
embedding of the fibers $\C\P^1\hookrightarrow Z(M)$ is holomorphic,
while the projection $Z(M)\to M$ is only a smooth map. We will use the notation 
$S(M):=\bS(\cS^+(M))$ for the unit sphere bundle of the spinor bundle.
The twistor and spinor bundles 
fit into the analog of diagram \eqref{diagHopfCP3},
\begin{equation}\label{HopfSZ}
\xymatrix{ S^1 \rto^{=} \dto & S^1 \dto &  \\ 
S^3 \rto \dto & S(M) \rto \dto & M \dto^{=} \\ \C\P^1 \rto & Z(M) \rto & M }
\end{equation}
The horizontal fibration comes from the identification $Z(M)=\bS(\Lambda_+(M))$ 
and the vertical one from $Z(M)=\P(\cS^+(M))$. We will also consider the associated diagram
\begin{equation}\label{CHopfSZ}
\xymatrix{ \C^* \rto^{=} \dto & \C^* \dto &  \\ 
\C^2\smallsetminus \{0 \} \rto \dto & \cS^+(M)^0 \rto \dto & M \dto^{=} \\ \C\P^1 \rto & Z(M) \rto & M }
\end{equation}
where $\cS^+(M)^0$ is the complement of the zero section in the spinor bundle.

\smallskip
\subsection{$\theta$-deformations and toric deformations}\label{thetadefSec}

We discuss our first noncommutative deformation method.
The Connes--Landi $\theta$-deformation method \cite{CoLa}, \cite{Yama} associates to a compact Riemannian spin
manifold $(X,g)$ that admits an action of a torus $T^2=U(1)\times U(1)$ by
isometries a noncommutative space $X_\theta$ with $\theta\in \R$
a deformation parameter. Here the notion of noncommutative space is
understood in the sense of spectral triples \cite{Co-S3}, a natural setting for
a noncommutative formulation of Riemannian spin geometry. 
In this setting, the original commutative manifold $(X,g)$ is encoded as
the data $(\cC^\infty(X), L^2(X,\bS), \Dirac)$ of its algebra of smooth
functions, the Hilbert space of square-integrable spinors, and the Dirac
operator. A reconstruction theorem \cite{Co-rec} shows that, conversely, 
a commutative spectral triple satisfying the relevant list of axioms
determines a classical manifold. Given a torus action by isometries
$T^2\hookrightarrow {\rm Isom}(X,g)$, the algebra $\cC^\infty(X)$ can
be deformed to a noncommutative algebra, which we denote by
$\cC^\infty(X)_\theta$, obtained by decomposing smooth functions in the
original algebra into Fourier modes (weighted components) with respect to the
torus action, and replacing their commutative pointwise product by a noncommutative
product modelled on the noncommutative $2$-torus $T^2_\theta$ of modulus $\theta$.
More precisely, by viewing functions $f\in \cC^\infty(X)$ as bounded multiplication operators 
on the Hilbert space $L^2(X,\bS)$, one decomposes $f$ into components $f_{n,m}$
according to the torus action, $\alpha_{(t_1,t_2)}(f_{n,m})=e^{2\pi i (n t_1+m t_2)} \, f_{n,m}$.
The deformed product of $\cC^\infty(X_\theta)$ is then defined 
component-wise by setting
\begin{equation}\label{thetadefprod}
 f_{n,m}\star_\theta h_{k.r}=e^{\pi i \theta (nr-mk)} f_{n,m} \, h_{k,r}. 
\end{equation} 
The Hilbert space and the Dirac operator of the spectral triple remain undeformed,
so that one obtains an isospectral deformation $X_\theta:=(\cC^\infty(X)_\theta, L^2(X,\bS), \Dirac)$.
A reconstruction theorem for theta-deformations is proved in \cite{Cac}.

\smallskip

For our main application here we are especially interested in the case of the $\theta$ deformation
of $S^3$ obtained by deforming all the tori $T^2$ in the Hopf fibration to noncommutative tori $T^2_\theta$.
This means that in the Hopf fibration $S^1\hookrightarrow S^3 \to S^2$ one considers the action of $T^2$
by translations on each of the tori of the Hopf foliation of $S^3$, translating the Hopf coordinates $(\xi_1,\xi_2)$. 
The effect of the $\theta$-deformation then transforms each $T^2$ in the foliation of $S^3$ with a 
noncommutative $T^2_\theta$ while maintaining the Hopf link given by the fibers over $0$ and $\infty$
undeformed. We refer to the resulting noncommutative space as $S^3_\theta$. 

\smallskip

More explicitly, we represent the sphere $S^3$ as the unit quaternions $q\in \H$
$$ q=\begin{pmatrix} z_1 & z_2 \\ - \bar z_2 & \bar z_1 \end{pmatrix} =\begin{pmatrix} e^{i\xi_1} \cos \eta & e^{i\xi_2} \sin \eta \\ - e ^{-i \xi_2} \sin \eta & e^{-i\xi_1} \cos \eta \end{pmatrix}, $$
with $z_1,z_2\in \C$, $|z_1|^2+ |z_2|^2=1$, and with $(\xi_1,\xi_2,\eta)$ the Hopf coordinates
$$ z_1 =x_1+ix_2= e^{i\xi_1} \cos\eta, \ \ \  z_2=x_3+i x_0=e^{i\xi_2}\sin\eta. $$
The $\theta$-deformation of the $3$-sphere replaces $q$ with 
$$ \begin{pmatrix} U \cos \eta & V \sin \eta \\ - V^* \sin \eta & U^* \cos \eta \end{pmatrix} $$
where $U,V$ are the generators of the noncommutative torus $T^2_\theta$ 
algebra, satisfying $UV=e^{2\pi i \theta} VU$. Thus, as shown in \cite{CoLa},
the algebra describing the noncommutative space $S^3_\theta$ is generated by $\alpha=U \cos \eta$
and $\beta =V \sin \eta$, satisfying the relations 
\begin{equation}\label{relsS3theta}
\alpha \beta = e^{2\pi i \theta} \beta\alpha, \ \ \ \alpha^* \beta = e^{-2\pi i \theta} \beta \alpha^*, \ \ \  \alpha^*\alpha=\alpha \alpha^*, \ \ \
\beta^*\beta =\beta\beta^*, \ \ \  \alpha \alpha^* + \beta \beta^*=1 . 
\end{equation}

\smallskip

We show that quantizing the $3$-sphere through the $\theta$-deformation that renders
all the Hopf tori noncommutative has the effect of generating a noncommutative
deformation of the sphere bundle of the spinor bundle of a self-dual $4$-manifold $M$, which
however leaves the twistor space $Z(M)$ classical. 

\begin{prop}\label{thetaSdef}
The $\theta$-deformation $S^3_\theta$ of the $3$-sphere determines a noncommutative deformation
$S(M)_\theta$ of the sphere bundle of the spinor bundle $S(M)=\bS(\cS^+(M))$ with the property that 
the noncommutative $S(M)_\theta$ fits into a diagram of fibrations
\begin{equation}\label{thetaHopfSZ}
\xymatrix{ S^1 \rto^{=} \dto & S^1 \dto &  \\ 
S^3_\theta \rto \dto & S(M)_\theta \rto \dto & M \dto^{=} \\ \C\P^1 \rto & Z(M) \rto & M }
\end{equation}
 where only the spaces $S^3_\theta$ and $S(M)_\theta$ are noncommutative and all the other
spaces, including the twistor space $Z(M)$, remain classical.
\end{prop}

\proof
If one considers the $\theta$-deformation $S^3_\theta$ of the $3$-sphere
considered above, one still has the Hopf fibration, where the total space $S^3_\theta$
is non-commutative, but both the base $S^2$ and the fiber $S^1$ remain commutative.
To see this, consider the $U(1)$-action on the algebra defining the noncommutative space
$S^3_\theta$ given by $\alpha \mapsto \lambda \alpha$ and $\beta \mapsto \lambda^{-1} \beta$,
for $\lambda\in U(1)$. This clearly preserves the defining relations. The invariant subalgebra
$(S^3_\theta)^{U(1)}$, which corresponds to the base of the fibration, is generated by the
elements $X=\beta\alpha$, $X^*=\alpha^*\beta^*$ and $Y=\alpha \alpha^* -\frac{1}{2}$
with the relations $XY=YX$, $YX^* = X^* Y$ and $Y^2+X X^* =\frac{1}{4}$, hence it is
the algebra of functions of a commutative $2$-dimensional sphere.
This Hopf fibration $S^1\hookrightarrow S^3_\theta \to S^2$ is considered from the
point of view of spectral triples and Dirac operators in \cite{DaSiZu}.

\smallskip

We then construct the deformation $S(M)_\theta$ by considering the fibration over the
commutative manifold $M$, where all the fibers are obtained by replacing the commutative
sphere $S^3$ with its $\theta$-deformation $S^3_\theta$. The resulting $S(M)_\theta$
can itself be regarded as a $\theta$-deformation, where the isometric action of $T^2$ 
on $S(M)$ used for the deformation is the action that translates the Hopf tori in each
fiber $S^3$. The defining algebra of $S(M)_\theta$ is generated by sections $\alpha(x), \beta(x)$
with $x\in M$, with the relations as in the case of $S^3_\theta$. 
The same argument used to show that the Hopf fibration $S^1\hookrightarrow S^3\to S^2$
becomes the fibration $S^1\hookrightarrow S^3_\theta \to S^2$ then shows that the
invariant subalgebra of the $U(1)$ action $\alpha(x)\mapsto \lambda \alpha(x)$ and
$\beta(x) \mapsto \lambda^{-1} \beta(x)$ has fibers over $M$ given by the quotient
$2$-spheres and is identified with the bundle $S^2\hookrightarrow Z(M)\to M$. Thus,
we obtain the fibration $S^1 \hookrightarrow S(M)_\theta \to Z(M)$ that fits the diagram above.
\endproof

\smallskip

This Connes-Landi $\theta$-deformation has an associated Cirio--Landi--Szabo toric
deformation, which is obtained by considering the diagram \eqref{CHopfSZ}. The main
idea behind this class of toric deformations is to deform algebraic tori $\bG_m(\C)^n=(\C^*)^n$
to noncommutative algebraic tori (as defined in \S 2.1 of \cite{CiLaSza1}) rather than 
deforming tori $T^n=(S^1)^n$ to the usual noncommutative tori. 

\smallskip

\begin{cor}\label{thetatoric}
There is a toric deformation $(\C^2\smallsetminus \{ 0 \})_\theta$ that fits into a Hopf fibration
$\C^*\hookrightarrow (\C^2\smallsetminus \{ 0 \})_\theta \to \C\P^1$. This determines a
corresponding noncommutative deformation of the diagram \eqref{CHopfSZ}.
\end{cor}

\proof The noncommutative toric deformation of $(\C^2\smallsetminus \{ 0 \})_\theta$ is
obtained by considering the toric structure and the deformation $\C[\sigma]_\theta$ 
of the algebras of the cones with the algebra $(\C^2\smallsetminus \{ 0 \})_\theta$ determined 
by a gluing diagram
$$ 0 \to (\C^2\smallsetminus \{ 0 \})_\theta \to \prod_\sigma \C[\sigma]_\theta \to 
\prod_{\sigma,\sigma'} \C[\sigma\cap \sigma']_\theta . $$
The gluing diagram is well defined because the algebras $\C[\sigma]_\theta$ are
subalgebras of the same noncommutative deformation of the ring of Laurent
polynomials associated to the maximal torus and the algebraic torus actions all agree. 
The explicit form of the relations that determine the maps $\C[\sigma]_\theta \to \C[\sigma\cap \sigma']_\theta$
is given in \cite{CiLaSza1}, p.54. The diagonal action of $\C^*$ on $\C^2\smallsetminus \{ 0 \}$ determines
a $\C^*$-action on the deformed algebra $(\C^2\smallsetminus \{ 0 \})_\theta$ with invariant subalgebra
that determines a commutative $\C\P^1$ so that the deformation fits into a Hopf fibration
$\C^*\hookrightarrow (\C^2\smallsetminus \{ 0 \})_\theta \to \C\P^1$. We can then consider the
noncommutative space obtained from $\cS^+(M)^0$ by deforming the $\C^2\smallsetminus \{ 0 \}$
fibers to $(\C^2\smallsetminus \{ 0 \})_\theta$ while leaving $M$ commutative, namely a bundle of
$(\C^2\smallsetminus \{ 0 \})_\theta$ algebras over $M$. The noncommutative space $\cS^+(M)^0_\theta$
obtained in this way fibers over the commutative twistor space $Z(M)$ with fibers $\C^*$.
\endproof

\smallskip

\subsection{Deformation quantization of the Hopf fibration and twistor spaces}\label{defqNCtwistSec}

We now consider a second type of noncommutative deformation of the
diagram \eqref{HopfSZ}, still based on a noncommutative deformation
of the Hopf fibration $S^1\hookrightarrow S^3 \to S^2$. This time,
however, both the base $S^2$ and the total space $S^3$ are deformed
to noncommutative spaces. We show that this deformation of the
Hopf fibration gives rise to compatible noncommutative deformations
of both the sphere bundle $S(M)$ of the spinor bundle and the twistor space $Z(M)$.
This method is based on deformation quantization. 
We show that the noncommutative twistor space $Z(M)_\hbar$ constructed
in this way differs from the quantization prescription of \cite{Pen68} through the
presence of one additional nontrivial commutator $[Z^\alpha, \overline{Z^\alpha}]=\hbar$.

\smallskip

It is well known that there are difficulties in applying the formalism of
deformation quantization to fibrations and principal bundles, including
the Hopf fibration $S^1\hookrightarrow S^3 \to S^2$ that we are
interested in here. This is discussed in detail in \cite{BNWW} (see also
Remark~2.10 of \cite{OMYM2}). In fact, a satisfactory very general theory 
of Riemannian principal bundles in noncommutative geometry was only 
developed very recently, \cite{BraBra}. In the next example we do not consider 
this more sophisticated viewpoint, as we work only at the level of the algebras,
not of spectral triples. 
We use here the construction of \cite{OMYM1}, \cite{OMYM2}, based
on a deformation quantization of contact manifolds. This allows us
to consider a noncommutative version of the Hopf fibration 
$S^1\hookrightarrow S^3 \to S^2$ that gives rise to compatible
deformation quantizations of $S^3$ and $S^2$. The latter
can be identified as a noncommutative K\"ahler manifold
deformation of $\C\P^1$, so that the complex manifold structure
of the twistor space is maintained. 

\smallskip

The Wick algebra is the algebra $\cA(\C^2_\hbar)$ of the quantum plane $\C^2_\hbar$, with generators 
$\zeta_0,\zeta_1,\zeta_0^\dagger, \zeta_1^\dagger$ (corresponding to the
two complex coordinates $\zeta_i$ and their conjugates $\bar\zeta_i$) 
and commutation relations
\begin{equation}\label{zetairels}
 [\zeta_i,\zeta_j]=0, \ \ \  [\zeta_i^\dagger, \zeta_j^\dagger]=0, \ \ \  [\zeta_i, \zeta_j^\dagger] = \hbar \delta_{ij}.
\end{equation} 
This algebra can be identified (see \cite{OMYM1}, \cite{OMYM2}) 
with a dense subalgebra of the deformation quantization 
$(\cC^\infty(\C^2)[[\hbar]], \star)$, with the associative product written in the Moyal form as
\begin{equation}\label{Moyalprod}
\begin{array}{l} 
f_1\star f_2 := f_1 \, \displaystyle{\exp (\hbar (\stackrel{\leftarrow}{\partial_\zeta}\stackrel{\rightarrow}{\partial_{\bar\zeta}} - 
\stackrel{\leftarrow}{\partial_{\bar\zeta}}\stackrel{\rightarrow}{\partial_\zeta} ))} \, f_2, \ \ \ \text{ where } \\[3mm]
f_1 \displaystyle{(\stackrel{\leftarrow}{\partial_\zeta}\stackrel{\rightarrow}{\partial_{\bar\zeta}} - 
\stackrel{\leftarrow}{\partial_{\bar\zeta}}\stackrel{\rightarrow}{\partial_\zeta} )} f_2 :=\displaystyle{\frac{1}{2} \sum_i (\partial_{\zeta_i} f_1 
\partial_{\bar\zeta_i} f_2 - \partial_{\bar\zeta_i}f_1 \partial_{\zeta_i} f_2} ),
\end{array}
\end{equation}
and all the terms in the expansion are bidifferential operators. 
Our notation here differs slightly from \cite{OMYM1}, \cite{OMYM2}, where the 
commutation relation of the Wick algebra is 
$[\xi_i, \bar \xi_j] =-2 \hbar \delta_{ij}$. Thus, our generators $\zeta_i, \zeta_i^\dagger$  of the algebra
are related to the generators $\xi_i, \bar \xi_i$ of \cite{OMYM1}, \cite{OMYM2} by $\zeta_i =\sqrt{2}\, \bar\xi_i$
and $\zeta_i^\dagger = \sqrt{2}\, \xi_i$. We work here with the version
as in \eqref{zetairels} for consistency with the commutation relations of the twistor
coordinates in \cite{Pen68}.

\smallskip

We now show that this noncommutative $\C^2_\hbar$,
with an associated noncommutative Hopf fibration, determine compatible 
quantizations of the sphere bundle $S(M)$ and the 
twistor spaces $Z(M)$. 

\smallskip

\begin{prop}\label{propNCtwistor}
The Wick algebra deformation $\cA(\C^2_\hbar)$ determines noncommutative
deformation quantizations of $S^3$ and $S^2$ compatible with the Hopf
fibration. Given a (anti-)self-dual $4$-manifold $M$ with twistor space $Z(M)$,
the deformation above induces compatible deformation quantizations of
$S(M)$ and $Z(M)$, related by a diagram 
\begin{equation}\label{qHopfSZ}
\xymatrix{ S^1 \rto^{=} \dto & S^1 \dto &  \\ 
S^3_\hbar \rto \dto & S(M)_\hbar \rto \dto & M \dto^{=} \\ S^2_\hbar \rto & Z(M)_\hbar \rto & M }
\end{equation}
\end{prop}

\proof
The relation of the Wick algebra to the Hopf fibration was described in detail in 
\cite{OMYM1}, \cite{OMYM2}. We recall the main steps here as we need them
in the construction of the noncommutative twistor space. 

\smallskip

In the Wick algebra consider the element 
$R^2:= \zeta_0^\dagger \star \zeta_0 + \zeta_1 \star \zeta_1^\dagger =\zeta_0 \star \zeta_0^\dagger + \zeta_1^\dagger \star \zeta_1$. 
In the algebra $(\cC^\infty(\C^2\smallsetminus \{ 0 \})[[\hbar]],\star)$ this is an invertible element
with a square root $R$ (Lemma~3.1, Theorem~3.3, and p.~929 of \cite{OMYM2}). 
Consider then the subalgebra $\cA$ generated by the elements
$\eta_i:= \sqrt{2}\, R^{-1} \star \zeta_i^\dagger$ and $\eta_i^\dagger:=\sqrt{2}\, \zeta_i \star R^{-1}$, and 
$\mu:=-2 \hbar R^{-2}$. 
These satisfy the relations
\begin{equation}\label{2sphererels}
\begin{array}{r}
[\mu^{-1}, \eta_i]=-\eta_i, \ \  [\mu^{-1}, \eta_i^\dagger]=\eta_i^\dagger, \ \  [\eta_0,\eta_1]=0, \\[2mm]
\eta_i\star \eta_j^* - (1-\mu) \eta_j^\dagger\star \eta_i =\mu \delta_{ij}, \ \  \eta_0^\dagger\star \eta_0+\eta_1^\dagger \star \eta_1 =1 . \end{array}
\end{equation}
As shown in \cite{OMYM1}, \cite{OMYM2}, the algebra $\cA=:\cA(S^3_\hbar)$ 
is a dense subalgebra of a closed subalgebra
$\cA^\infty$ of $(\cC^\infty(\C^2\smallsetminus \{ 0 \})[[\hbar]],\star)$ which is isomorphic to a deformation 
quantization of the $3$-sphere, $\cA^\infty\simeq \cC^\infty(S^3)[[\mu]]$. The algebra $\cA^\infty$ can
also be characterized as the fixed point subalgebra of the flow $\rho: \R \to {\rm Aut}( \cC^\infty(\C^2\smallsetminus \{ 0 \})[[\hbar]], \star)$ with $\rho_t \zeta_i = e^t \zeta_i$ and $\rho_t \zeta_i^\dagger=e^t \zeta_i^\dagger$, and with
$\rho_t\hbar =e^{2t}\hbar$.  The quantized $2$-sphere at the
base of the Hopf fibration is obtained by considering the algebras $\cC^\infty(U_i)[[\hbar]]$ with
$U_i\subset \C\smallsetminus \{ 0 \}$ given by $\{ \zeta_i \neq 0 \}$. The algebra $\cA^\infty$
admits localizations $\cA^\infty_{U_i}$, which are the invariant subalgebras, under the flow $\rho_t$
of $\cC^\infty(U_i)[[\hbar]]$, and the algebra $\cA^\infty$ is obtained as a gluing
$$ \xymatrix{ & \cA^\infty \dlto \drto & \\ \cA^\infty_{U_0} \drto & & \cA^\infty_{U_1} \dlto \\
& \cA^\infty_{U_0\cap U_1} & } $$
The elements $Z=\zeta_0^{-1} \star \zeta_1=\eta_0^{-1}\star \eta_1$ and $W=\zeta_1^{-1}\star \zeta_0$
are defined on $U_0$ and $U_1$, respectively, and satisfy $[\mu,Z]=[\mu,Z^\dagger]=[\mu,W]=[\mu,W^\dagger]=0$.
Thus, one can consider the subalgebras of $\cA^\infty_{U_i}$ generated, respectively, by 
$\mu,Z,Z^\dagger,\eta_0,\eta_0^\dagger$ and $\mu,W,W^\dagger,\eta_1,\eta_1^\dagger$, with the transition
function on $U_0\cap U_1$ given by $(\mu,W,W^\dagger,\eta_1,\eta_1^\dagger)=(\mu,Z^{-1},(Z^\dagger)^{-1},Z\star \eta_0, \eta_0^\dagger\star Z^\dagger)$. Here $Z$ and $W$ define the local coordinates on 
the deformed $\C\P^1$ and $\eta_0,\eta_1$, which satisfy $[Z, \eta_0]=[W,\eta_1]=0$ and $\eta_1=Z\star \eta_0$, 
can be regarded as holomorphic sections of a line bundle over this deformed $\C\P^1$. The coordinates
$Z,Z^\dagger$ satisfy $[Z,Z^\dagger]=\mu (1+Z\star Z^\dagger)\star (1+Z\star Z^\dagger)$, which is regarded
in \cite{OMYM2} as a deformation of the K\"ahler metric on $\C\P^1$ satisfying $\{ z, \bar z \}=(1+ z \bar z)^2$.
The canonical conjugate variable of $Z$ in this deformed $\C\P^1$ is not $Z^\dagger$ but
$(1+Z^\dagger\star Z)^{-1} \star Z^\dagger$, since these satisfy the commutation relation
$[ Z, (1+Z^\dagger\star Z)^{-1} \star Z^\dagger ] =\mu$, see Lemma~4.2 of \cite{OMYM2}. 
Thus, we obtain in this way a consistent deformation quantization of $S^3$ and $\C\P^1$.
The respective deformation parameters in this construction of \cite{OMYM1}, \cite{OMYM2}
are related by $\mu =-2 \hbar R^{-2}$, where $\hbar$ is the deformation parameter of $\C^2$ in the 
Wick algebra $\C^2_\hbar$ and $\mu$ the deformation parameter of $S^3$ and $\C\P^1$. For
simplicity, we will refer to all three of these deformations using the same notation 
$\C^2_\hbar$, $S^3_\hbar$, and $\C\P^1_\hbar$, and the corresponding algebras as $\cA(\C^2_\hbar)$,
$\cA(S^3_\hbar)$, and $\cA(\C\P^1_\hbar)$, where the dependence of the deformation
parameter $\mu$ on $\hbar$ is as stated above. 

\smallskip

Consider then a self-dual Riemannian $4$-manifold $M$ with its twistor space
$Z(M)=\bS(\Lambda_+(M))=\P(\cS^+(M))$ and with the sphere bundle 
$S(M)=\bS(\cS^+(M))$ of the spinor bundle. Let $\cU=\{ U_\alpha \}$ be
an open covering of $M$ that trivializes both $S(M)$ and $Z(M)$. We denote
by $\phi^Z_{\alpha\beta}$ and $\phi^S_{\alpha\beta}$ a set of transition functions
for $Z(M)$ and $S(M)$, respectively, compatible with the Hopf fibration
$S^1\hookrightarrow S^3 \to S^2$ in the diagram \eqref{HopfSZ},
$$ \xymatrix{ S^1 \rto^{id} \dto & S^1 \dto \\ S^3 \rto^{\phi^S_{\alpha\beta}} \dto & S^3 \dto \\
S^2 \rto^{\phi^Z_{\alpha\beta}} & S^2 } $$
where $\phi^S_{\alpha\beta}$ are the $SU(2)$-valued transition functions of the spinor bundle
$\cS^+(M)$ acting on the associated sphere bundle $\bS(\cS^+(M))$. We construct then a bundle
of Wick algebras over $M$, seen as a deformation of the spinor bundle $S^+(M)$. Namely,
we consider over each open set $U_\alpha$ the trivial product $U_\alpha\times \C^2_\hbar$
which means the algebra given by the tensor product $\cA(U_\alpha)\otimes \cA(\C^2_\hbar)$
of the Wick algebra with functions on $U_\alpha$. The the $SU(2)$-valued transition 
functions of the spinor bundle $\phi^S_{\alpha,\beta}$ determine algebra automorphisms 
$\phi^S_{\alpha,\beta}: U_\alpha \cap U_\beta \to {\rm Aut}(\C^2_\hbar)$ which act by
mapping the generators $(\zeta_0,\zeta_1)\mapsto (\tilde\zeta_0=a\zeta_0+b \zeta_1, \tilde\zeta_1=c\zeta_0+d\zeta_1)$ and $(\zeta_0^\dagger,\zeta_1^\dagger)\mapsto (\tilde\zeta_0^\dagger=\bar a\zeta_0^\dagger+\bar b \zeta_1^\dagger, \tilde\zeta_1^\dagger=\bar c\zeta_0^\dagger+\bar d\zeta_1^\dagger)$.
The $\tilde\zeta_0,\tilde\zeta_1,\tilde\zeta_0^\dagger,\tilde\zeta_0^\dagger$ are generators of
the Wick algebra with $[\tilde\zeta_i,\tilde\zeta_j]=0$, $\tilde\zeta_i^\dagger,\tilde\zeta_j^\dagger]=0$ and
$[\tilde\zeta_i,\tilde\zeta_j^\dagger]=(|a|^2+|b|^2) \hbar \delta_{ij}=\hbar \delta_{ij}$. 
Thus, identifying $\cA(U_\alpha \cap U_\beta)\otimes \cA(\C^2_\hbar)$ with the algebra $\cA(\cU_\alpha\cap U_\beta, \cA(\C^2_\hbar))$ of $\cA(\C^2_\hbar)$-functions on $U_\alpha \cap U_\beta$ we obtain automorphisms
$\Phi_{\alpha,\beta}(F)(x)=\phi^S_{\alpha,\beta}(x) (F(x))$.
We denote by $\cS^+(M)_\hbar$ the nocommutative space obtained in this way through
the gluing with the transition functions described above,
$$ 0 \to \cA(\cS^+(M)_\hbar) \to \prod_\alpha \cA(U_\alpha)\otimes \cA(\C^2_\hbar) \to \prod_{\alpha,\beta} 
\cA(U_\alpha \cap U_\beta)\otimes \cA(\C^2_\hbar). $$
We obtain a noncommutative $S(M)_\hbar=\bS(\cS^+(M))_\hbar$ with the same construction, with fiber
the algebra $\cA(S^3_\hbar)$. Indeed, the algebra $\cA^\infty\simeq \cC^\infty(S^3)[[\mu]]$ is characterized
as the subalgebra of $\cC^\infty(\C\smallsetminus \{ 0 \})[[\hbar]]$ invariant under the flow $\rho_t$
and the change of variables $(\zeta_0,\zeta_1,\zeta_0^\dagger,\zeta_1^\dagger)\mapsto (\tilde\zeta_0,\tilde\zeta_1,\tilde\zeta_0^\dagger,\tilde\zeta_1^\dagger)$ commutes with the flow.
We then obtain the compatible noncommutative
twistor space $Z(M)_\hbar$ by considering, over the same open covering, a locally trivial bundle 
of algebras $\C\P^1_\hbar$ with the transition functions induced by the transition functions $\phi^S_{\alpha,\beta}$.
The algebra of $\C\P^1_\hbar$ is characterized in \cite{OMYM2}
as the subalgebra of $\cA^\infty$ given by the condition $\{ f\in \cA^\infty\,|\, [\mu,f]=0 \}$. This depends on
the fact that $[\mu, Z]=[\mu,W]=0$, which in turn is determined by the commutator relations of the
element $R^2$ with the generators of the Wick algebra, which with the commutation relations 
\eqref{zetairels} is given by $\zeta_i \star R^2 = (R^2 + \hbar) \zeta_i$. Thus, it suffices 
to check that the $SU(2)$ action $(\zeta_0,\zeta_1)\mapsto (\tilde\zeta_0,\tilde\zeta_1)$ preserves
these commutation relations. This is the case since the $\tilde\zeta_i$ are linear combinations of the $\zeta_i$.
Thus, we obtain compatible comstructions of the noncommutative $S(M)_\hbar$ and $Z(M)_\hbar$ as
bundles of noncommutative algebras over the commutative space $M$, that fit the diagram \eqref{qHopfSZ}.
\endproof

\smallskip

We now compare the noncommutative twistor space $Z(M)_\hbar$ obtained in this
way with the noncommutative twistor space introduced by one of us in \cite{Pen68}
and we show that our $Z(M)_\hbar$ has one additional nontrivial commutator relation.

\begin{prop}\label{propNCtwistor2}
Let $M$ be a self-dual $4$-manifold and let $Z(M)_\hbar$ be the noncommutative twistor space
obtained as in Proposition~\ref{propNCtwistor}. The quantized $S(M)_\hbar$ and $Z(M)_\hbar$ differ 
from the quantization prescription of \cite{Pen68} by the presence of two rather than one nontrivial
commutators, $[Z^\alpha, \bar Z_\alpha]=\hbar$ and $[Z^\alpha, \overline{Z^\alpha}]=\hbar$.
\end{prop}

\proof
In twistor coordinates $Z^\alpha$ and $\bar Z_\alpha$, the classical variables $\bar Z_\alpha$ 
are the twistor conjugate variables, namely
$\bar Z_0=\overline{Z^2}$, $\bar Z_1=\overline{Z^3}$, $\bar Z_2=\overline{Z^0}$, $\bar Z_3=\overline{Z^1}$.
The quantization of twistor space introduced in \cite{Pen68} is obtained by imposing 
the condition that the variables $Z^\alpha$ commute with each other, $[Z^\alpha, Z^\beta]=0$, 
and also  $[\bar Z_\alpha, \bar Z_\beta]=0$, while $Z^\alpha$ and $\bar Z_\alpha$
are conjugate variables satisfying the relation $[Z^\alpha, \bar Z_\beta]=\hbar \delta^\alpha_\beta$.
The $\hbar$ parameter can be absorbed into a rescaling of the variables, 
but we will consider it here explicitly as deformation parameter, to compare with the Wick
algebras considered above. 
In the case of $M=S^4=\H\P^1$ we can consider the two affine charts of $\H\P^1$ given by
$(q,1)$ and $(1,\tilde q)$ with $q,\tilde q\in \H$ and the transition function  $\tilde q=q^{-1}$
on the overlap $\H\smallsetminus \{ 0 \}$. We write $q=z_0+z_1 j$ with $z_0,z_1\in \C$.
For the construction of $\cS^+(S^4)_\hbar$ as in
Proposition~\ref{propNCtwistor}, we then consider an algebra of the form 
$\cA(\C^2)\otimes\cA(\C^2_\hbar)$ for each of these two open charts. 
These have generators $z_i\otimes \zeta_j$ for $i,j\in \{0,1\}$, satisfying the relations
$[z_i \otimes \zeta_j, z_a\otimes \zeta_b]=0$, $[\bar z_i \otimes \zeta_j^\dagger, \bar z_a\otimes \zeta_b^\dagger]=0$
and $[z_0\otimes\zeta_i, \bar z_1\otimes \zeta_i^\dagger]=\hbar$. Thus,
we obtain an identification of $\cS^+(S^4)_\hbar$ with the algebra with generators
$Z^\alpha =z_\alpha \otimes \zeta_\alpha$ and $\bar Z_\alpha =\bar z_\alpha \otimes \zeta_\alpha^\dagger$
with commutation relations $[Z^\alpha, Z^\beta]=0=[\bar Z_\alpha, \bar Z_\beta]$ and with two 
nontrivial commutator relations between the $Z^\alpha$ and the $\bar Z_\alpha$ given by
$[Z^\alpha, \bar Z_\beta]= \hbar$ when either $\alpha =\beta$ or $\alpha + 2 =\beta$ mod $4$. 
In terms of the conventional notation with twistor variables mentioned above this means both
$[Z^\alpha, \bar Z_\alpha]= \hbar$ and $[Z^\alpha, \overline{Z^\alpha}]= \hbar$.
\endproof

\smallskip


\smallskip
\subsection{Fuzzy twistor spaces}\label{fuzzySec}

We mention one more possible
construction that satisfies a form of the Hopf fibration compatibility and the
commutativity of spacetime and which also is close to satisfying \eqref{relQtwistor}.
This is related to the fuzzy sphere approximations of the $2$-sphere.

\smallskip

Consider first the noncommutative deformation $\C^2_\hbar$, with 
generators $\zeta_0,\zeta_1,\zeta_0^\dagger, \zeta_1^\dagger$
(corresponding to the usual coordinates $z_0,z_1,\bar z_0, \bar z_1$
of the commutative case) and with the commutation relations \eqref{zetairels}.
For a single copy of $\C$ the deformation $\C_\hbar$ with $[\zeta,\zeta^\dagger]= \hbar$
just corresponds to the quantum plane where the real coordinates satisfy $[y,x]=2i \hbar$ and
as an algebra $\C^2_\hbar=\C_\hbar \otimes \C_\hbar$ is a product of two such quantum planes,
which we will also write as $\C_\hbar \times \C_\hbar$.

\smallskip

In this setting, the noncommutative deformation of the Hopf fibration is obtained by
considering the Wick algebra $\cA(\C^2_\hbar)$ generated by the 
$\zeta_0,\zeta_1,\zeta_0^\dagger, \zeta_1^\dagger$ satisfy the commutation 
relations \eqref{zetairels} and the algebras $\cA((\C^2\smallsetminus \{ 0 \})_\hbar)$
and $\cA(S^3_\hbar)$ as described in \S \ref{defqNCtwistSec}.

\smallskip

Consider then the elements 
\begin{equation}\label{La}
L_1 =\frac{1}{2}(\zeta_0 \zeta_1^\dagger + \zeta_0^\dagger \zeta_1), \ \ \  L_2 = \frac{i}{2}(\zeta_0 \zeta_1^\dagger - \zeta_0^\dagger \zeta_1), \ \ \  L_3 = \frac{1}{2}(\zeta_0^\dagger \zeta_0 - \zeta_1^\dagger \zeta_1).
\end{equation}
These generate a $U(1)$-invariant subalgebra for the action 
$\zeta_i\mapsto \lambda \zeta_i$ and
$\zeta_i^\dagger \mapsto \bar\lambda \zeta_i^\dagger$ 
and they satisfy the commutation relation
$[L_a,L_b]=i \hbar \epsilon^{abc} L_c$. By regarding the subalgebra 
generated by the $L_a$ as a deformed $2$-sphere, one can view
the inclusion of this subalgebra as a version of a deformed Hopf fibration. 

\smallskip

Thus, we see that the noncommutative twistor spaces $Z(M)_\hbar$, obtained via deformation quantization,
have an associated family of ``fuzzy twistor spaces" based on the relation between the
deformed $2$-sphere described here above and the fuzzy spheres $S^2_N$. In turn the
fizzu spheres have a direct connection with deformation quantization of the $2$-sphere, 
as discussed in \cite{FreiKras}.
The fuzzy spheres \cite{Mad} of level $N=2j$ determine an approximation of
the ordinary $2$-sphere $S^2$ by finite noncommutative spaces.
These are based on decomposing the algebra of functions on the
$2$-sphere, seen as a $\cU({\mathfrak su}(2))$-module, into irreducible
representations $\oplus_{\ell\geq 0}\cV_\ell$, with $\cV_\ell$ spanned
by the spherical harmonics $\Theta_{\ell,m}$, and then truncating at some
energy level $N=2j$, by only considering $0\leq \ell\leq 2j$.
A description of the fuzzy spheres in terms of spectral triples is given in \cite{DaLiVa} and a
precise sense in which the fuzzy spheres converge to the ordinary sphere when $N\to\infty$
is analyzed in \cite{Rief}. 

\smallskip

The fuzzy sphere algebra $S^2_N$ is obtained by mapping the coordinates $(x_1,x_2,x_3)$ 
of the $2$-sphere $S^2\subset \R^3$, with $x_1^2+x_2^2+x_3^2 =1$, to operators 
$$ X_a := \frac{1}{\sqrt{j(j+1)}} \, J_a, $$
where $N=2j$ and $J_a$ the generators of the Lie algebra ${\mathfrak su}(2)$ satisfying
$[J_a,J_b]=i \epsilon^{abc} J_c$, viewed as operators acting in the $(N+1)$-dimensional representation of $SU(2)$.
The normalization factor is chosen so that the sphere relation $\sum_a X_a^2 =1$
is preserved. The map $x_a \mapsto X_a$ is not an algebra homomorphism, but it
determines an isomorphism of $\star$-representations of $\cU({\mathfrak su}(2))$.
The resulting algebra $S^2_N$ describing the fuzzy sphere is generated by 
$X_1,X_2,X_3$ with the relation $X_1^2+X_2^2+X_3^2=1$ and the nontrivial commutation relation
\begin{equation}\label{fuzzyrel}
 [ X_a, X_b ]=\frac{1}{\sqrt{j(j+1)}} i \epsilon_{abc} X_c. 
\end{equation} 
This algebra is in fact just the matrix algebra $M_{N+1}(\C)$. 

\smallskip

Under the map $x_a \mapsto X_a$ the
spherical harmonics $\Theta_{\ell,m}$ are mapped to matrices $\hat\Theta_{\ell,m}\in M_{N+1}(\C)$ 
(the fuzzy spherical harmonics), whose entries are Clebsch--Gordon coefficients, and where
one retains only the harmonics with $\ell=0,\ldots,N$.
Thus, the algebra $S^2_N$ can be equivalently described by considering the expansion
$f(x)=\sum_{\ell,m} a_{\ell,m} \Theta_{\ell,m}(x)$ in spherical harmonics $\Theta_{\ell,m}$
of functions on the $2$-sphere and replacing $f(x)$ with the element 
$\hat f =\sum_{\ell=0}^N \sum_m a_{\ell,m} \hat\Theta_{\ell,m}$ in $S^2_N$. 
For functions on $S^2$ that only involve modes in the spherical harmonics with $\ell\leq N$
the fuzzy sphere product is then given by $f_1 \star_{S^2_N} f_2:=\hat f_1 \cdot \hat f_2$
as product of the corresponding matrices in $M_{N+1}(\C)$. 
As shown in \cite{FreiKras}, this product of the fuzzy sphere algebra $S^2_N$ is related to
the deformation quantization product of $S^2_\hbar$ by the relation
$$ f_1 \star_{S^2_N} f_2  = \cP_N ( f_1 \star_\hbar f_2 )|_{\hbar=2/(N+1)}, $$
where $\cP_N$ denotes the projection of the first $N+1$ modes $\ell=0,\ldots,N$
in the spherical harmonics and $f_1,f_2$ are in the range of $\cP_N$. 

\smallskip

The construction of fuzzy twistor spaces is similar to the construction of the
noncommutative twistor spaces based on deformation quantization discussed
in \S \ref{defqNCtwistSec}. 

\begin{prop}\label{fuzzyTwistor}
The fuzzy sphere algebra $\cA(S^2_N)=M_{N+1}(\C)$ seen as a subalgebra of the
Wick algebra $\C^2_\hbar$ for $\hbar=1/\sqrt{j(j+1)}$ and $N=2j$, determines
fuzzy twistor spaces $Z(M)_N$ and $S(M)_N$ compatible with the Hopf fibration
\eqref{HopfSZ}.
\end{prop}

\proof
The fuzzy twistor spaces $Z(M)_N$ are obtained
by considering an open covering $\{ U_\alpha \}$ of $M$
that trivializes the spinor bundle $\cS^+(M)$, with $SU(2)$-valued
partition functions $\phi^S_{\alpha,\beta}$. We then consider
over each $U_\alpha$ the algebra $\cA(U_\alpha)\otimes \C^2_\hbar$
and the associated noncommutative space $\cS^+(M)_\hbar$ obtained
by gluing these algebras with the transition functions as in Proposition~\ref{propNCtwistor}.
For  $\hbar=1/\sqrt{j(j+1)}$ and $N=2j$, the 
fuzzy twistor space is then obtained by considering the subalgebras
$\cA(U_\alpha)\otimes \cA(S^2_N)= \cA(U_\alpha)\otimes M_{2j+1}(\C)$
with the transition functions $\phi^S_{\alpha,\beta}$ acting as automorphisms
of the algebra $\cA(S^2_N)$ using the $(N+1)$-dimensional representation of $SU(2)$. 
\endproof

\smallskip

As mentioned above, for $\hbar=1/\sqrt{j(j+1)}$ and $N=2j$, the algebra describing the 
fuzzy sphere $S^2_N$ is the matrix algebra $M_{2j+1}(\C)$, hence the fuzzy twistor 
space $Z(M)_N$ is an almost-commutative geometry in the sense of \cite{Cac2}. 
We will discuss some of the properties of this almost-commutative geometry 
more in detail in the next section. 

\smallskip
\subsection{Geometric quantization of twistor spaces and the Hopf fibration}\label{PenroseQHopfSec}

We return now to the original geometric quantization of twistor
spaces \cite{Pen68}, recalled in \S \ref{PenroseQSec} above 
and we discuss the role of the Hopf fibration \eqref{HopfSZ}, in
comparison with the other cases introduced above.

\smallskip

In the Riemannian setting, the Hopf fibration is involved in the geometric quantization of the
twistor space in the form of the $\C^*$-bundle $\cS^+(M)_0$ over
the twistor space $Z(M)$ and the complex structure $J$ on
$T(\cS^+(M)_0)$ compatible with the complex structure on $Z(M)$,
with the twistor coordinates $Z^\alpha$ and $\bar Z_\alpha$.
However, in this case, the role of the Hopf fibration is
more subtle than in the other forms of quantization we described in
this section. 

\smallskip

One can see this by focusing on the
Riemannian case with $M=S^4$, with $Z(M)=\C\P^3$ and $S(M)=S^7$.
In this case we can see explicitly that if the commutation prescription \eqref{relQtwistor}
is obtained as a Wick algebra deformation and we also impose  
the same compatibility requirements with the Hopf fibration 
diagram \eqref{HopfSZ} used in the previous constructions,  
that would necessarily lead to a noncommutative $S^4$. 

\smallskip

For $M=S^4$ we have $Z(M)=\C\P^3$ and $S(M)=S^7$ and these
spaces fit in the diagram \eqref{diagHopfCP3} of Hopf fibrations, or
equivalently in the diagram \eqref{diagCHopfCP3}. We now require
that $\C^4$ is quantized as a Wick algebra $\C^4_\hbar$, with generators 
$\zeta_0,\zeta_1,\zeta_2,\zeta_3$ and
$\zeta_0^\dagger ,\zeta_1^\dagger,\zeta_2^\dagger,\zeta_3^\dagger$
and commutation relations $[\zeta_i,\zeta_j]=0$,  $[\zeta_i^\dagger,\zeta_j^\dagger]=0$
and $[\zeta_i, \zeta^\dagger_j]=\hbar \delta_{ij}$. This agrees with the
commutators \eqref{relQtwistor} for the noncommutative twistor
space of $S^4$ by identifying the variables $Z^\alpha$ with the generators $\zeta_i$
and the variables $\bar Z_\alpha$ with the generators $\zeta_i^\dagger$. 
We also require that the resulting quantizations of $Z(S^4)=\C\P^3$ and of
$S(S^4)=S^7$ are compatible with the Hopf fibration diagrams 
\eqref{diagHopfCP3}  and \eqref{diagCHopfCP3}. This means that
the prescription for the quantization of $(\C^4\smallsetminus \{ 0 \})_\hbar$
should be compatible with the projection maps $\C^4\smallsetminus \{ 0 \} \to \H\P^1$
and $S^7\to \H\P^1$. 

\smallskip

The commutative algebra $\cA(S^4)$ of functions on $S^4$ has commuting
generators $\alpha, \alpha^\dagger, \beta, \beta^\dagger, x$ with relation
$\alpha \alpha^\dagger + \beta \beta^\dagger + x^2 =1$ and the projection
map is given by (see Appendix A of \cite{LaWvS})
$$ \alpha = 2 (z_0 \bar z_2 + z_1 \bar z_3), \ \ \  \beta= 2(z_1z_2 - z_0 z_3), \ \ \ 
x= z_0 \bar z_0 + z_1 \bar z_1 - z_2 \bar z_2 - z_3 \bar z_3 . $$
The subalgebra of $(\C^4\smallsetminus \{ 0 \})_\hbar$
generated by the elements
$$ \alpha := 2 (\zeta_0 \zeta_2^\dagger + \zeta_1 \zeta_3^\dagger), \ \ \  \beta := 2( \zeta_1 \zeta_2 - \zeta_0 \zeta_3), \ \ \  x= \zeta_0 \zeta_0^\dagger + \zeta_1 \zeta_1^\dagger - \zeta_2 \zeta_2^\dagger - \zeta_3 \zeta_3^\dagger $$
satisfies
$$ \frac{1}{4} [\alpha, \beta] =  \zeta_0 \zeta_1 (\zeta_2^\dagger  \zeta_2 -  \zeta_2  \zeta_2^\dagger )
- \zeta_0 \zeta_1 (\zeta_3^\dagger \zeta_3 -  \zeta_3 \zeta_3^\dagger ) = 0 $$
$$ \frac{1}{4} [\alpha, \alpha^\dagger] = \zeta_0 \zeta_0^\dagger \zeta_2^\dagger  \zeta_2 -
\zeta_0^\dagger \zeta_0 \zeta_2  \zeta_2^\dagger +  \zeta_1  \zeta_1^\dagger \zeta_3^\dagger  \zeta_3
- \zeta_1^\dagger \zeta_1 \zeta_3  \zeta_3^\dagger $$
$$ = \hbar \zeta_2^\dagger  \zeta_2 -\hbar \zeta_0^\dagger \zeta_0 + \hbar \zeta_3^\dagger  \zeta_3
- \hbar \zeta_1^\dagger \zeta_1 = - \hbar \, x  = \hbar\, (R_1^2  - R_0^2) $$ 
$$ \frac{1}{4} [\beta, \beta^\dagger] = \zeta_1 \zeta_1^\dagger \zeta_2 \zeta_2^\dagger - \zeta_1^\dagger  \zeta_1 \zeta_2^\dagger \zeta_2 + \zeta_0  \zeta_0^\dagger \zeta_3 \zeta_3^\dagger - \zeta_0^\dagger 
\zeta_0 \zeta_3^\dagger\zeta_3 $$
$$ = \hbar \zeta_2 \zeta_2^\dagger + \hbar \zeta_1^\dagger  \zeta_1 +\hbar \zeta_3 \zeta_3^\dagger +\hbar 
\zeta_0^\dagger \zeta_0 = \hbar (R_0^2 + R_1^2) , $$
where as before we write $R_0^2=\zeta_0^\dagger \zeta_0 + \zeta_1 \zeta_1^\dagger$ and $R_1^2 = \zeta_2^\dagger \zeta_2 +\zeta_3 \zeta_3^\dagger$. 
$$ \frac{1}{4} [\alpha, \beta^\dagger] = - [\zeta_0,\zeta_0^\dagger] \zeta_2^\dagger \zeta_3^\dagger + [\zeta_1, \zeta_1^\dagger] \zeta_2^\dagger \zeta_3^\dagger =0 $$
and $[\beta,\alpha^\dagger]=0$ likewise. 
The commutators $[\alpha,\alpha^\dagger]$ and $[\beta,\beta^\dagger]$ do not simultaneously vanish,
hence the subalgebra obtained in this way is also noncommutative.

\smallskip

Moreover, we can see that, if we adapt to the Hopf fibration $S^3\hookrightarrow S^7 \to S^4$
the argument used in \S \ref{defqNCtwistSec} for the deformation quantization of the Hopf
fibration $S^1\hookrightarrow S^3 \to S^2$, by replacing complex numbers with quaternions,
we also end up with a noncommutative $\H\P^1_\hbar$ obtained as the noncommutative
$\C\P^1_\hbar$ discussed in \S \ref{defqNCtwistSec}. In this case also the commutator
relations \eqref{relQtwistor} are satisfied and the same strict compatibility with the Hopf fibration used
in our other constructions of quantized twistor spaces are 
satisfied, but this cannot be made compatible with the requirement that the spacetime
manifold $S^4$ remains commutative. 

\smallskip

The discussion above shows that the compatibility between the quantization
of twistor spaces by commutators \eqref{relQtwistor} and the Hopf fibration is not
implemented by the  diagrams \eqref{diagHopfCP3} and
\eqref{HopfSZ}. However, a different form of compatibility with the Hopf fibration holds
for the twistor quantization of \cite{Pen68}. 
In order to better identify the role of the Hopf fibration in the geometric quantization of
the twistor space of \cite{Pen68}, it is useful to look at the construction in the original
setting of a Lorentzian metric, and the occurrence of the Hopf fibration of $S^3$ 
(Clifford parallels) in the Lorentzian version of twistor theory. 

\smallskip

\begin{prop}\label{PenHopfDiag}
The geometric quantization of (Lorentzian) twistor spaces of \cite{Pen68} with commutator relations \eqref{relQtwistor}
is compatible with the Hopf fibration, by viewing copies of the Hopf fibration 
$S^1 \hookrightarrow S^3 \to S^2$ embedded in the Hopf fibration
$\C\P^1 \hookrightarrow \C\P^3 \to S^4$ by first restricing the projection $\C\P^3 \to S^4$
to $\P N \to S^3$ over the equatorial sphere of $S^4$ and then slicing $\P N$ with
planes $P$ in $\C\P^3$ passing through a chosen point $q$ in the upper half $\P T^+$ of
$\C\P^3 \smallsetminus \P N$.
\end{prop}

\proof
Here, in the Lorentzian case, $\C\P^3$ has
an $SU(2,2)$ rather than an $SU(4)$ structure. The subspace $\P N$ of $\C\P^3$,
consisting of the element of zero $SU(2,2)$ norm, divides $\C\P^3$ into two halves $\P T^\pm$,
respectively of positive and negative norms. The $S^2$ fibration of $\C\P^3$ over 
$S^4$ has $\P N$ over an $S^3$ equatorial subspace of the sphere $S^4$.
Now, to see the Hopf fibrations, we take an arbitrary point $q$ in the top (positive norm) 
half of $\C\P^3$ and take an arbitrary $\C\P^2$ plane $P$ through $q$, of the kind which 
contains a projective line in the bottom half of $\C\P^3$, so that $P$ has positive $SU(2,2)$ norm. 
We find that the intersection of $P$ with $\P N$ is a Hopf-fibred $S^3$, where the Hopf circles 
are the intersections of the projective lines through $q$ with this $S^3$. 
The $S^2$ fibration of $\C\P^3$ carries these Hopf fibrations down to the equatorial $S^3$ in the 
sphere $S^4$.

\smallskip

The $S^4$ here is not really ``physical space-time", but it may be thought of 
as having the physical $3$-space at time $t=0$, represented by the equatorial $S^3$, but where the 
$S^4$ arises when the time $t$ evolves away from zero through pure-imaginary numbers 
(a so-called ``Wick rotation"). Thus, $S^4$ should be regarded as 
the conformally compactified Wick-rotated space-time.

\smallskip

The original importance of these Hopf fibrations to Lorentzian twistor theory (and whence the 
original name ``twistor") came about from the fact that the points of $\P N$ have an immediate 
physical interpretation, in terms of light rays in the Minkowskian space-time. The way that we 
can ``see" the points in $\C\P^3$ (or, more directly, the planes $P$ in $\C\P^3$) in physical terms, 
is in terms of these twisted congruences of light rays in the physical $3$-space, here represented 
as the equatorial $S^3$ described above.

\smallskip

This means that the diagram illustrating the role of the Hopf fibration in the geometric
quantization of twistor space of \cite{Pen68} is not the one we considered in \eqref{diagHopfCP3}
but it arises by considering the inclusions
\begin{equation}\label{Hopfinclusions}
 \xymatrix{ \C\P^1 \ar@{^{(}->}[r] & \C\P^3 \ar[r] & S^4 \\
\C\P^1 \ar[u]^{=} \ar@{^{(}->}[r] & \P N \ar@{^{(}->}[u] \ar[r] & S^3 \ar@{^{(}->}[u] \\
S^1 \ar@{^{(}->}[u] \ar@{^{(}->}[r]  & S^3 \ar@{^{(}->}[u] \ar[r] & S^2 \ar@{^{(}->}[u] } 
\end{equation}
where the second line is obtained by restricting the fibration $\C\P^1\hookrightarrow \C\P^3 \to S^4$
over the equatorial $S^3$ in $S^4$ and the third line is obtained by slicing $\P N$ with a plane
$P=\C\P^2$ through a chosen point $q$ in the upper (positive norm) half of $\C\P^3$, and 
correspondingly slicing the fibration of $\P N$ over $S^3$.

\smallskip

The submanifold $\P N$ of the twistor space $\C\P^3$ is a level set $\P N =\{ K=\sum_\alpha Z^\alpha \bar Z_\alpha =0 \}$ of the signature $(+,+,-,-)$ norm associated to the $SU(2,2)$ structure on $\C\P^3$ mentioned above. 
The symplectic form $\omega =\sum_\alpha dZ^\alpha \wedge d\bar Z_\alpha$ on the twistor space satisfies
$\omega =i \partial \bar \partial K= d( dK \circ J)$. The $1$-form $\alpha=dK \circ J |_{\P N}$ determines the contact structure on $\P N$ with distribution of contact hyperplanes $\xi={\rm Ker}(\alpha) = T \P N \cap J\, T\P N$, with $J$ the complex structure. The geometric quantization of the twistor space $\C\P^3$ as a symplectic manifold, 
that we recalled in  Section~\ref{PenroseQSec}, induces a compatible quantization of the contact manifold $\P N$. (For a general formalism for geometric quantization of symplectic manifolds with contact boundary see for example \cite{WeiZam}.) The planes $P$ through a point $p\in \P T^+$ are symplectic submanifolds and 
$T (P\cap \P N) \cap \xi$ determines the associated distribution of contact planes on the Hopf spheres
$S^3 =P\cap \P N$. One obtains in this way a quantization of the Hopf fibration
$S^1\hookrightarrow S^3 \to S^2$ compatible with the geometric quantization of the twistor space.
\endproof

\smallskip
\subsection{Another $\theta$-deformation} \label{SecondThetaSec}

The role of the Hopf fibration $S^1\hookrightarrow S^3 \to S^2$ in the case of Lorentzian twistor spaces,
described in \eqref{Hopfinclusions} and Proposition~\ref{PenHopfDiag} suggests then a different use of
the Connes--Landi $\theta$-deformations to obtain a noncommutative deformation of twistor spaces.
Instead of deforming $S^3$ to the noncommutative $S^3_\theta$ in the diagrams \eqref{diagHopfCP3}
and \eqref{HopfSZ}, as we discussed in Proposition~\ref{thetaSdef}, which gives a noncommutative $S(M)_\theta$
with commutative $Z(M)$, we can apply the same $\theta$-deformation of the Hopf fibration, with
noncommutative $S^3_\theta$ and commutative $S^1$ and $S^2$, to all the Hopf spheres
$S^3=P \cap \P N$ in \eqref{Hopfinclusions}. This gives rise to a resulting $\theta$-deformation for the
twistor space $Z(M)=\C\P^3$, or of more general twistor spaces in the Lorentzian case. Notice that,
while the spacetime manifold is Lorentzian, and Lorentzian geometry is explicitly used to identify
copies of the Hopf fibration $S^1\hookrightarrow S^3 \to S^2$ inside  the Hopf fibration
$\C\P^1 \hookrightarrow Z(M) \to M$, only the Riemannian structure of $S^3$ is used in these
$\theta$-deformations as the Lorentzian spacetime manifold remains undeformed and classical.
Thus, the formalism of $\theta$-deformations (which requires the Riemannian setting of spectral
triples) can still be applied.  We summarise this reasoning with the following statement, whose
proof is analogous to Proposition~\ref{thetaSdef}.

\smallskip

\begin{prop}\label{otherThetaDef}
A noncommutative Connes--Landi $\theta$-deformation of the twistor space $Z(M)=\C\P^3$
and of the Hopf fibration $\C\P^1\hookrightarrow \C\P^3_\theta \to S^4$ can be
obtained by simultaneously applying the Connes--Landi $\theta$-deformation 
$S^1 \hookrightarrow S^3_\theta \to S^2$ to all the Hopf spheres $S^3 = P\cap \P N$ 
with $P$ varying over planes in $\C\P^3$ passing through a given point $p\in \P T^+$.
\end{prop}

\smallskip

There is a significant difference between a noncommutative deformation of twistor
space obtained as in Proposition~\ref{otherThetaDef} and the geometric quantization of 
\cite{Pen68}. In the case discussed here, the noncommutative deformation is entirely 
carried by the Hopf spheres $S^3_\theta$ that deform the intersections $P\cap \P N =S^3$.
Thus, the noncommutativity only affects the $\P N$ part of twistor space rather than the
entire $\P T^\pm$ parts. Significant examples of classical spaces with noncommutative
boundaries occur elsewhere, for example the noncommutative boundaries of modular
curves studied in \cite{ManMar}. On the other hand, in the quantization of \cite{Pen68}
it is the entire twistor space that is quantized through its symplectic structure, with a
compatible quantization of the contact submanifold $\P N$.

\section{Deformations and Gluing}\label{DefGlueSec}

In this section we consider the problem of gluing noncommutative twistor spaces
formulated in \cite{Penrose2020}, and we present a general setting based on
the Gerstenhaber--Schack complex, \cite{GeSch}, to address this question for
noncommutative twistor spaces obtained via a procedure of deformation quantization.

\smallskip

In the commutative
case, for (anti)self-dual Riemannian manifolds, the gluing of twistor spaces that
corresponds to a connected sum of spacetime manifolds is analyzed in \cite{DoFr}
in terms of Kodaira--Spencer deformation theory, in the form developed in \cite{Fri} for
singular spaces with normal crossings singularities, applied to the gluing along
the exceptional divisors of the blowups of the twistor spaces along one of the twistor lines.
Here we consider the more general deformation theory, as formulated in \cite{GeSch},
which involves both commutative and noncommutative deformation. We formulate the
problem of gluing quantized twistor spaces in terms of the deformation theory of a 
diagram of algebras. We first discuss the general setting and then we apply it to
the different forms of quantization of twistor spaces illustrated in the previous section.

\smallskip
\subsection{Noncommutative deformation and obstructions}

The construction of \cite{DoFr} of the gluing of twistor spaces corresponding to connected sums of the
underlying self-dual $4$-manifolds relies essentially on the Kodaira--Spencer deformation theory for
complex manifolds. Since in our setting we are dealing with noncommutative twistor spaces, we first
recall here a setting, the Gerstenhaber--Schack complex, where the usual Kodaira--Spencer deformation 
theory can be recovered as part of a more general deformation theory of diagrams of unital associative
algebras. We follow the exposition of \cite{GeSch} for this summary. For our purposes we restrict to
the case of algebras over $\C$, though the setting of \cite{GeSch} is much more general. 

\smallskip

In this setting, the deformation theory for a single unital associative algebra $\cA$ over $\C$ is governed
by its Hochschild cohomology $\oplus_n HH^n(\cA,\cA)$.  
Consider a deformation of an associative algebra $\cA$, namely a 
$\C[[t]]$-algebra $\cA[[t]]$ with $\C[[t]]$-linear multiplication 
$\alpha_t=\alpha + t\alpha_1 + t^2 \alpha_2 + \cdots$ extending the multiplication $\alpha$ of $\cA$ with 
$\C$-linear maps $\alpha_i : \cA\times \cA \to \cA$, satisfying the associativity 
relation $\alpha_t(\alpha_t(a,b),c)=\alpha_t(a,\alpha_t(b,c))$. 
Using the notation $f \star g (a,b,c):=f(g(a,b),c)-f(a,g(b,c))$, we can rewrite the associativity
constraint as a sequence of equations 
\begin{equation}\label{associativityn}
\sum_{p+q=n, p,q>0} \alpha_p \star \alpha_q(a,b,c) = a \alpha_n(b,c)-\alpha_n(ab,c)
+\alpha_n(a,bc) - \alpha_n(a,b) c, 
\end{equation}
where the right-hand-side is the Hochschild coboundary $\delta \alpha_n (a,b,c)$.
The first order term $\alpha_1$ satisfies  $\delta \alpha_1=0$, 
so that $\alpha_1$ defines a Hochschild 2-cocycle and cocycles that
differ by a coboundary determine equivalent deformations. Thus, one identifies $HH^2(\cA,\cA)$ as
parameterizing the infinitesimal deformations of $\cA$. The next condition gives
$\alpha_1(\alpha_1(a,b),c) -\alpha_1(a,\alpha_1(b,c)) =\delta \alpha_2$, where the left-hand-side defines
a Hochschilf $3$-cocycle. Thus, this constraints represents a possible obstruction to extending the
infinitesimal deformation $\alpha_1$ to a global deformation $\alpha_t$. One can view this as
a quadratic map $\Theta: HH^2(\cA,\cA) \to HH^3(\cA,\cA)$. The condition $\Theta([\alpha_1])=0$
is the necessary vanishing of the primary obstruction that corresponds to the second associativity constraint. 
Similarly, the expressions $\sum_{p+q=n, p,q>0} \alpha_p \star \alpha_q$ define $3$-cocycles and the
constraints \eqref{associativityn} require that all of these are coboundaries $\delta \alpha_n$, hence
trivial in $HH^3(\cA,\cA)$. 

\smallskip

The deformation theory of a single associative algebra $\cA$ is generalized in \cite{GeSch} to 
a deformation theory of diagrams of algebras. In this setting, given a small category $\cC$, a diagram
of associative $\C$-algebras over $\cC$ is a contravariant functor $\bA: \cC^{op} \to {\rm Alg}_\C$. 
Examples include the cases where $\cC$ is the poset of open sets of a smooth manifold ordered
by inclusion or the poset of Stein open sets of a complex manifold, seen as a category, and 
associated commutative algebras of smooth or holomorphic functions, respectively, 
with restriction maps. The formal deformations of a diagram $\bA$ are diagrams of $\C[[t]]$-algebras
over the same $\cC$ that reduce modulo $t$ to $\bA$. This means, for every object $C$ of $\cC$
a deformed associative multiplication $\alpha_t^C= \alpha^C + t \alpha^C_1+ t^2 \alpha^C_2 + \cdots$ on
the corresponding algebra $\bA^C$ and for every morphism $\phi: C \to C'$ a $\C[[t]]$-algebra
morphism $\phi_\bA: \bA_t^{C'} \to \bA^C_t$ of the deformed algebras, so that one obtains a
diagram $\bA_t: \cC^{op} \to {\rm Alg}_{\C[[t]]}$. A suitable notion of equivalence of diagrams
and deformations is discussed in \S 17 of \cite{GeSch}. A single algebra $\cA=\cA_\bA$ can be associated
to a diagram $\bA: \cC \to {\rm Alg}_\C$.  It is defined as a convolution product over the
diagram in the following way. As a $\C$-vector space $\cA$ is spanned by elements of the form
$\sum a^C \, \phi_{\bA,C}$, with 
elements $a^C \in \bA^C$ for objects $C\in {\rm Obj}(\cC)$ and morphisms
$\phi_{\bA,C}\in {\rm Mor}(\cC)$ with source $s(\phi_{\bA,C})=C$. The convolution product is 
determined on the individual components by
 \begin{equation}\label{algdiagram}
 (a^C\, \phi_{\bA,C}) \cdot (a^{C'} \, \psi_{\bA,C'}) = a^C \phi_{\bA,C} (a^{C'}) (\phi\psi)_{\bA,C} 
 \end{equation}
 when $t(\phi)=s(\psi)=C'$ and zero otherwise, with $(\phi\psi)_{\bA,C}=\psi_{\bA,C'} \circ \phi_{\bA,C}$.
 The deformation theory of diagrams $\bA$ is constructed in \S 21 of \cite{GeSch} in terms
 of a cochain complex that computes a generalization of a local cohomology for a local
 system over the nerve of the category $\cC$, which is given by a Yoneda cohomology.
    
 \smallskip
 
 The subdivision $\cC'$ of a small category $\cC$ is a category whose
 simplicial nerve $\cN(\cC')$ is the first barycentric subdivision of the nerve
 $\cN(\cC)$.  The second subdivision $\cC''$ is always a poset. The subdivision 
 comes endowed with a functor $\cC'\to \cC$, hence a diagram
 $\bA: \cC \to {\rm Alg}_\C$ has an associated subdivision $\bA':\cC'\to {\rm Alg}_\C$
 by precomposition. We denote by $\cA'$ and $\cA''$ the assembled algebras
 associated to the subdivisions $\bA'$ and $\bA''$ of the diagram. One also denotes
 by $\bA_\#$ the extension of the diagram $\bA$ to the category $\cC_\#$ where a
 terminator object $\infty$ has been added with a unique map $C \to \infty$ from
 every $C\in {\rm Obj}(\cC)$, by setting $\bA^\infty=\C$ and $\C \to \bA^C$ the
 unique homomorphism determined by the $\C$-algebra structure of $\bA^C$. 
  In the case where the small category $\cC$ is a poset, there is an isomorphism of
 Hochschild homologies $HH^*(\bA,\bA)\simeq HH^*(\cA,\cA)$ of a diagram $\bA$ of
 algebras and of its associated single algebra $\cA=\cA_\bA$. This is the ``special cohomology
 comparison theorem" of \cite{GS}. For more general small categories $\cC$ a similar
 ``general cohomology comparison theorem" holds (\cite{GeSch}, \S 23) which identifies
 the Hochschild homology $HH^*(\bA,\bA)\simeq HH^*(\cA'',\cA'')$. These identifications
 are then used (see \S 25 of \cite{GeSch}) to compare the deformation theory of the
 diagram $\bA$ with that of the assembled algebras $\cA$ and $\cA''$. Indeed, it is
 proved in \S 25 of \cite{GeSch} that the deformation theory of a diagram $\bA$ is
 equivalent to the deformation theory of the single algebra $\cA''_\#$ (see p.~224 of \cite{GeSch}).
 
 \smallskip
 
 In particular, in the case of a complex manifold $X$, with $\cT=\cT_X$ the
 sheaf of (germs of) holomorphic tangent vector fields, and with $\Lambda^k \cT$ the 
 exterior powers, a covering of $X$ consisting of Stein open sets (or affine open sets in the
 case of projective algebraic varieties) closed under intersections
 determines a poset $\cC$ and a diagram $\bA: \cC\to {\rm Alg}_\C$ of commutative algebras 
 with $\bA^U$ the ring of holomorphic functions on the open set $U$. Then the Hochschild homology of the
 diagram is given by $HH^n(\bA,\bA)=\oplus_{\ell +k=n} H^\ell (X, \Lambda^k \cT)$,
 where the terms $H^\ell(X,\Lambda^k \cT)$ are identified with the terms 
 $HH^{\ell, k}(\bA,\bA)$ defined more generally for a diagram of commutative algebras 
 in \S 26 of \cite{GeSch}. These identifications $HH^{\ell, k}(\bA,\bA)\simeq H^\ell (X, \Lambda^k \cT)$
 were proved in \cite{GeSch} for the case where $X$ is a smooth projective variety and $\cC$ the
 poset determined by a covering of affine open sets and conjectured for the case of a complex
 manifold with Stein open sets. A more general setting where these identifications hold, which
 includes complex analytic manifolds and smooth schemes in characteristic zero is given 
 in \cite{Schu}. In particular, all the infinitesimal deformations of the diagram $\bA$ of
 commutative algebras are parameterized by $HH^2(\bA,\bA)$, with an obstruction map
 $HH^2(\bA,\bA) \to HH^3(\bA,\bA)$. Among these deformations, the part 
 $HH^{1,1}(\bA,\bA)\simeq H^1(X,\cT)$ parameterizes deformations of $\bA$ to
 diagrams of commutative algebras, that is, classical deformations of the underlying manifold $X$,
 with the obstruction map $HH^{1,1}(\bA,\bA) \to H^{2,1}(\bA,\bA)$ identified with the classical
 obstruction map $H^1(X,\cT)\to H^2(X,\cT)$. These are deformations ``in the commutative direction".
 The part $HH^{2,0}(\bA,\bA)\simeq H^0(X,\Lambda^2\cT)$ of the space classifying infinitesimal deformations
 of $\bA$ corresponds instead to those deformations of the diagram of commutative algebras to diagrams
 of {\em non-commutative} associative algebras, deformations ``in the noncommutative direction".
 
 \smallskip
 \subsection{Classical and noncommutative deformations of twistor spaces}\label{DefSec} 
 
 We analyze here the classical and noncommutative deformation theory of
 the twistor spaces $Z_i=Z(M_i)$ of two (anti)self-dual Riemannian manifolds $M_i$
 and of their blowups $\tilde Z_i$ along a fixed twistor line. We describe the classical
 and noncommutative deformation theory of the gluing $\tilde Z$ of the blowups along
 the exceptional divisors in terms of the Hochschild cohomology of an associated
 diagram of algebras as in \cite{GeSch}.
We start by showing how to associate to the gluing 
$\tilde Z(M)=\tilde Z(M_1) \sqcup_{E_1\simeq E_2} \tilde Z(M_2)$,
of the blowups $\tilde Z(M_i)={\rm Bl}_{\C\P^1}(Z(M_i))$ along the exceptional divisors a
diagram of algebras in the sense of \cite{GeSch}.

\begin{lem}\label{tildeZdiagA}
The singular space $\tilde Z(M)$ obtained by gluing the blowups $\tilde Z(M_i)$ along the
exceptional divisors $E_i$ determines an associated diagram of commutative algebras
$\bA(\tilde Z): \cC \to {\rm Alg}_\C$, where $\cC$ is a poset determined by a system of
Stein open sets in the complement of $E_i$ in $\tilde Z_i$ and pairs of Stein open sets in 
$\tilde Z_1$ and $\tilde Z_2$ that contain the identified exceptional divisors.
\end{lem}

\proof
Let $\gamma: T_{x_1}(M_1) \to T_{x_2}(M_2)$ be the orientation reversing isometry of the
tangent spaces of the spacetime manifolds $M_i$ at the points $x_i$ where the connected
sum is performed. We denote by the same symbol $\gamma$ the induced identification of
the exceptional divisors $\gamma: E_1 \to E_2$ of the blowups of the twistor spaces $Z(M_i)$
at the twistor lines $\C\P^1_{x_i}$. 
Let $\cU_i = \{ U_{i,\alpha} \}$ be open coverings of the blown up twistor spaces $\tilde Z_i=\tilde Z(M_i)$
by Stein open sets, closed under intersections. They form a poset under inclusions.
Consider then the small category $\cC$ with objects given by those $U_{i,\alpha}$ in the
coverings $\cU_i$ with the property that $U_{i,\alpha}\cap E_i=\emptyset$ and
additional objects given by pairs $(U_{1,\alpha}, U_{2,\beta})$ of open sets in these
coverings such that $E_1 \cap U_{1,\alpha} \neq \emptyset$ and $E_2\cap U_{2,\beta}\neq \emptyset$
and such that $\gamma: E_1 \cap U_{1,\alpha} \to E_2\cap U_{2,\beta}$ is an isomorphism. Morphisms
of $\cC$ between open sets of each covering $\cU_i$ are inclusions and morphisms
between pairs $(U_{1,\alpha}, U_{2,\beta})$ and $(U_{1,\alpha'}, U_{2,\beta'})$ are pairs of
inclusions $\iota_{1,\alpha,\alpha'}: U_{1,\alpha} \hookrightarrow U_{1,\alpha'}$ and
$\iota_{2,\beta,\beta'}: U_{1,\beta} \hookrightarrow U_{1,\beta'}$ with the property that
$\iota_{2,\beta,\beta'}|_{E_2\cap U_{2,\beta}} \circ \gamma = \iota_{1,\alpha,\alpha'} |_{E_1 \cap U_{1,\alpha}}$. 
We then construct a functor $\bA: \cC \to {\rm Alg}_\C$ by assigning to objects $U_{i,\alpha}$
$\bA^{U_{i,\alpha}}=\cA(U_{i,\alpha})$ the algebra of holomorphic functions on $U_{i,\alpha}$
with morphisms $\bA(\iota_{i,\alpha,\alpha'})=\rho_{i,\alpha',\alpha}: \cA(U_{i,\alpha'}) \to \cA(U_{i,\alpha})$
the restriction map corresponding to the inclusion $\iota_{i,\alpha,\alpha'}: U_{i,\alpha} \hookrightarrow U_{i,\alpha'}$.
To objects $(U_{1,\alpha}, U_{2,\beta})$ we assign the algebra $\bA^{(U_{1,\alpha}, U_{2,\beta})}$ given by
$$ \{ (f_{1,\alpha}, f_{2,\beta})\,:\,
f_{1,\alpha} \in \cA(U_{1,\alpha}), \,   f_{2,\beta}\in \cA(U_{2,\beta}), \,\, f_{2,\beta}|_{E_2\cap U_{2,\beta}}\circ \gamma = f_{1,\alpha}|_{E_1\cap U_{1,\alpha}} \}, $$
with morphisms $(\iota_{1,\alpha,\alpha'},\iota_{2,\beta,\beta'})$ with $\iota_{2,\beta,\beta'}|_{E_2\cap U_{2,\beta}} \circ \gamma = \iota_{1,\alpha,\alpha'} |_{E_1\cap U_{1,\alpha}}$ in $\cC$ mapped to the restriction maps
$\rho_{\alpha',\beta', \alpha,\beta}: \bA^{(U_{1,\alpha'}, U_{2,\beta'})} \to \bA^{(U_{1,\alpha}, U_{2,\beta})}$. 
\endproof

\smallskip

The general construction of the assembled algebra $\cA_\bA$ associated to a diagram and
the special cohomology comparison theorem of \cite{GeSch} then give the following.

\begin{cor}\label{AlgtildeZdiag}
The deformation theory of the diagram $\bA(\tilde Z)$ is equivalent to the deformation
theory of a single algebra generated by elements of the form $f_{i,\alpha} \rho_{i,\alpha',\alpha}$
and $(f_{1,\alpha},f_{2,\beta})\, \rho_{\alpha',\beta', \alpha,\beta}$ with the convolution product
$$  f_{i,\alpha} \rho_{i,\alpha',\alpha} \cdot f_{i,\alpha'} \rho_{i,\alpha'',\alpha'} =
f_{i,\alpha} \rho_{i,\alpha',\alpha}(f_{i,\alpha'}) \, \rho_{i,\alpha'',\alpha} $$
$$  (f_{1,\alpha},f_{2,\beta})\, \rho_{\alpha',\beta', \alpha,\beta} \cdot (f_{1,\alpha'},f_{2,\beta'})\, \rho_{\alpha'',\beta'', \alpha',\beta'} = (f_{1,\alpha} \rho_{\alpha', \alpha}(f_{1,\alpha'}),f_{2,\beta} \rho_{\beta',\beta} (f_{2,\beta'}) )\, \rho_{\alpha'',\beta'', \alpha,\beta} $$
and zero otherwise.
\end{cor}

\smallskip

The computation of the Hochschild cohomology that governs the deformation theory of
the diagram $\bA(\tilde Z)$ then gives the following result that recovers the Donaldson--Friedman
deformation theory of the singular space $\tilde Z$ as the part of the deformation theory of
the diagram $\bA(\tilde Z)$ that corresponds to deformations ``in the commutative direction".

\begin{thm}\label{defAtildeZ}
The commutative part of the deformation theory of the diagram $\bA(\tilde Z)$ recovers the
Donaldson--Friedman deformation theory of the singular space $\tilde Z$.
\end{thm}

\proof Near the normal crossings singular locus, the 
gluing $\tilde Z=\tilde Z_1 \sqcup_{E_1\stackrel{\gamma}{\simeq} E_2} \tilde Z_2$
is locally described by $\{ z_0 z_1 =0 \} \subset \C^4$ and 
we can assume that the open coverings $\cU_i$ of $\tilde Z_i$ are chosen so that 
this local description holds for each $U_{1,\alpha}\sqcup_{U_{1,\alpha}\cap E_1\simeq E_2\cap 
U_{2,\beta}} U_{2,\beta}$. Thus, we can view the algebras $\cA_{\alpha,\beta}:=\cA(U_{1,\alpha}, U_{2,\beta})$
as copies of the algebra associated to $V=\{ z_0 z_1 =0 \} \subset \C^4$. In this case, the Hochschild cohomology
is computed by Andr\'e--Quillen cohomology, namely the decomposition 
$HH^n(\cA,\cA)=\oplus_r HH^{n-r,r}(\cA,\cA)$ satisfies $ HH^{n-r,r}(\cA,\cA)\simeq T^{n-r,r}(\cA)$, where
the Andr\'e--Quillen cohomology  $T^{i,j}(\cA)$ is the $j$-th cohomology group 
of $\Hom_\cA(\Lambda^i \bL_A, A)$ for the derived exterior power $\Lambda^i \bL_A$ 
of the cotangent complex, see \cite{Loday} \S 3.5.4. 
The terms $T^{i,1}(\cA)$ correspond to the terms defined as $\bT^i_V$ of \cite{DoFr} and
are identified with the piece $HH^{n-1,1}(\cA,\cA)=T^{n-1,1}(\cA)$ of $HH^n(\cA,\cA)$. In terms of
deformation theory, the term $HH^{1,1}(\cA,\cA)$ of the second Hochschild cohomology 
parameterizes the infinitesimal deformations in the ``commutative direction'', 
while the term $HH^{2,0}(\cA,\cA)$ of the second Hochschild cohomology represents the
infinitesimal deformations of $\cA$ in the ``noncommutative direction". The obstruction
map for the classical deformations is given by the component $\Phi: HH^{1,1}(\cA,\cA) \to HH^{2,1}(\cA,\cA)$
of the overall obstruction map $\Phi: HH^2 (\cA,\cA) \to HH^3(\cA,\cA)$. This corresponds to the
obstruction map $\Phi: \bT^1_V=T^{1,1}(\cA) \to \bT^2_V=T^{2,1}(\cA)$ considered in \cite{DoFr}. 
To see then that this identification holds not only at the local level of the algebras $\cA_{\alpha,
\beta}$ but also globally for $\tilde Z$, we can use the fact that the Hochschild cohomology
$HH^*(\bA,\bA)$ for diagrams of algebras and the pieces $HH^{r,n-r}(\bA,\bA)$ of the decomposition
can be computed in terms of two filtrations (that truncate the first rows or columns, respectively) on
a double complex $C^{*,*}(\bA,\bA)e(r)$, with $e(r)$ the idempotent that determines the $(r,n-r)$ piece,
which has the Hochschild differential on the vertical direction and the simplicial differential of the nerve
of the category $\cC$ in the horizontal direction and building blocks given by 
$\prod_{\dim \sigma =p} C^{*+r} (\bA^{c\sigma}, \bA^{c\sigma})e(r) \otimes_{\bA^{c\sigma}} \bA^{d\sigma}$
where a $p$-simplex $\sigma: [p]\to \cC$ is a covariant functor from the category $[p]=\{ 0< \cdots < p \}$
to $\cC$ and $c\sigma=\sigma(0)$ and $d\sigma=\sigma(p)$. These filtrations determine a spectral
sequence converging to $HH^*(\bA,\bA)$, see \S 21--26 of \cite{GeSch}. For $\bA(\tilde Z)$ one obtains
in this way the deformation theory $\Phi: \bT^1_{\tilde Z} \to \bT^2_{\tilde Z}$ of \cite{DoFr} from the component
$\Phi: HH^{1,1}(\bA(\tilde Z), \bA(\tilde Z)) \to HH^{2,1}(\bA(\tilde Z),\bA(\tilde Z))$ of the deformation
theory $\Phi: HH^2(\bA(\tilde Z),\bA(\tilde Z))\to HH^3(\bA(\tilde Z),\bA(\tilde Z))$ of the diagram
of algebras. 
\endproof

\smallskip

The result above shows, in particular, that even when the
deformation theory $\Phi: \bT^1_{\tilde Z} \to \bT^2_{\tilde Z}$ of \cite{DoFr}
for the gluing $\tilde Z$ of the blowups of the twistor spaces $Z(M_i)$ is
obstructed, it may still be possible to obtain an unobstructed deformation
theory in the ``noncommutative direction", that is, for the infinitesimal deformations in
$HH^{2,0}(\bA(\tilde Z),\bA(\tilde Z))$. This means that, in such cases, even if the
singular $\tilde Z$ cannot be deformed commutatively to the smooth twistor space
$Z(M)$ for the connected sum $M=M_1\# M_2$ (for instance if $M$ does not
admit a (anti)self-dual structure) one still has a noncommutative twistor space
$Z(M)_\hbar$ obtained as a deformation in the noncommutative direction
of $\tilde Z$. 

\smallskip

We also need a consistency relation between the choices of the noncommutative
deformations on the twistor spaces $Z_i=Z(M_i)$ and on the glued $\tilde Z$. 
This can be obtained by first showing that the deformations of the twistor spaces
$Z_i$ induce deformations of the blown up twistor spaces $\tilde Z_i$ and then
by identifying a compatibility condition between the deformations of the $\tilde Z_i$
and the deformation of $\tilde Z$.

\smallskip

We first recall the following setting. As shown in \cite{Block}, given a diagram of algebras
\begin{equation}\label{diagAAi}
 \xymatrix{ A \rto^{\phi_1} \dto^{\phi_2} & A_1 \dto^{\psi_1} \\ A_2 \rto^{\psi_2} & A_3 } 
\end{equation}
such that 
\begin{equation}\label{modseqAAi}
 0 \to A \stackrel{\phi_1\oplus \phi_2}{\to} A_1 \oplus A_2 \stackrel{\psi_1-\psi_2}{\to} A_3 \to 0 
\end{equation} 
is an exact sequence of $A$-bimodules, with the properties that the maps $A\to A_i$ are
flat epimorphisms, then there is a long Mayer--Vietoris exact sequence for Hochschild homology
$$ \cdots \to HH_n(A,A)\to HH_n(A_1,A_1)\oplus HH_n(A_2,A_2) \to HH_n(A_3,A_3)\to \cdots $$
We cannot apply this to directly to the case of the spaces $\tilde Z_i$ and $\tilde Z$ and their local models
near the normal crossings singularity $E_1\simeq E_2$, because the algebra homomorphisms
$\phi_i$ in the corresponding diagram do not satisfy the flatness hypothesis. 
Thus, we cannot compare directly the deformation classes
and the obstructions for $\tilde Z_i$ and $\tilde Z$ through the Mayer--Vietoris sequence.
However, there is still a long exact sequence of Hochschild cohomology that we can use to
compare these deformations.

\smallskip

Let $\cA_{\alpha,\beta}$ be one of the algebras describing the geometry of $\tilde Z$ near the
normal crossings singularity in the diagram of algebras $\bA(\tilde Z)$ and let $\cA_{\alpha,1}$
and $\cA_{\beta,2}$ be algebras in the diagrams $\bA(\tilde Z_1)$ and $\bA(\tilde Z_2)$, respectively,
describing the geometry near the exceptional divisor $E_i$. For simplicity of notation we drop the subscripts
$\alpha, \beta$ and we just refer to these algebras as $\cA, \cA_1, \cA_2$. 

\smallskip

\begin{lem}\label{MVseq}
Let $\gamma_i \in HH^2(\cA_i, \cA_i)$ for $i=1,2$ be unobstructed deformation classes, $\Phi(\gamma_i)=0\in HH^3(\cA_i,\cA_i)$, and let $\gamma \in HH^2(\cA,\cA)$ be a deformation class that is also unobstructed,
$\Phi(\gamma)=0 \in HH^3(\cA,\cA)$. There are epimorphisms $\phi_i: \cA \to \cA_i$ that induce a long exact
sequence of Hochschild cohomology
\begin{equation}\label{longAAi}
 \cdots \to HH^n(\cA,\cA) \to HH^n(\cA,\cA_1)\oplus HH^n(\cA,\cA_2) \to HH^n(\cA,\cA_3) \to 
  \cdots 
\end{equation} 
and morphisms $HH^n(\cA_i,\cA_i)\to HH^n(\cA,\cA_i)$. The image $c_i(\gamma)$ of $\gamma$ in 
$Z^2(\cA,\cA_i)$ is a $2$-cocycle that extends to a $2$-cocycle in the complex $(Z^2(\cA,\cA_i)[[t]], \delta)$
where $\delta c =\delta_i c + [m_\gamma, c]$ with $\delta_i$ the Hochschild differential of $C^*(\cA,M)$ for the
bimodule $M=\cA_i$ and $m_\gamma$ the deformed multiplication on $\cA$ determined by the unobstructed
deformation class $\gamma\in HH^2(\cA,\cA)$. Similarly, the image $c(\alpha_i)$ of $\gamma_i$ in
$Z^2(\cA,\cA_i)$ is a $2$-cocycle that extends to a $2$-cocycle in the complex $(Z^2(\cA,\cA_i)[[t]], \delta)$. 
\end{lem}

\proof In our case the geometry near the normal crossings singularity can be described as 
the locus $\{ z_1 z_2 =0 \}$ with $\{ z_1 =0 \}$ and $\{ z_2 =0 \}$ the two components 
and $\{ z_1=z_2=0 \}$ the intersection. The corresponding algebras $\cA,\cA_1,\cA_2,\cA_3$
then fit into a diagram \eqref{diagAAi} which satisfies the exactness of the associated sequence
of $\cA$-modules \eqref{modseqAAi} and the epimorphism condition, which can also be stated
as the condition that $\cA_i\otimes_\cA \cA_i\simeq \cA_i$. The short exact sequence \eqref{modseqAAi}
of $\cA$-modules induces a long exact sequence \eqref{longAAi} of Hochschild cohomology 
(see \cite{GeSch}, p.~36). Moreover, the Hochschild cohomology is a contravariant functor in the
algebra, hence the homomorphisms $\phi_i: \cA \to \cA_i$ 
induce homomorphisms $\phi_i^*: HH^n(\cA_i,\cA_i)\to HH^n(\cA,\cA_i)$.
Consider an unobstructed deformation class $\gamma \in HH^2(\cA,\cA)$. The condition
$\Phi(\gamma)=0\in HH^3(\cA,\cA)$, ensuring that all obstructions vanish, is the condition
that the left-hand-side of \eqref{associativityn}
are all coboundaries for all $n$. A homomorphism $\phi: M \to N$ of $\cA$-modules
induces a morphism $C^n(\cA,M)\to C^n(\cA,N)$ by composition, mapping
a multilinear map $f: \cA \times \cdots \times \cA \to M$ by to the multilinear map
$\phi \circ f: \cA \times \cdots \times \cA \to N$, which is a cochain map. We still denote
by $\gamma$ a $2$-cocycle representing the deformation of $\cA$ and by $\phi_i(\gamma)$
its image in $Z^2(\cA,\cA_i)$. The $3$-cochains $\Phi(\gamma)_n$ in the left-hand-side of \eqref{associativityn}
are similarly mapped to $3$-cochains $\phi_i \circ \Phi(\gamma)_n$ in $C^3(\cA,\cA_i)$.
We need to check that these cochains are the cochains that determine the extensibility
condition of the cocycle $\phi_i(\gamma)\in Z^2(\cA,\cA_i)$ to a $2$-cocycle in 
the complex $(Z^*(\cA,\cA_i)[[t]],\delta_\gamma)$. This extensibility condition is
discussed in \cite{Piper}. A $2$-cocycle $\upsilon \in Z^2(\cA,M)$
extends to a $2$-cocycle $\upsilon_t$ to $(Z^*(\cA,M)[[t]],\delta_\gamma)$ iff
$\upsilon_t$ is determined by a choice of a collection
$\upsilon_n \in C^2(\cA,M)$ satisfying the property that all the obstructions $3$-cochains
$$ \omega_n(\upsilon):= \sum_{p+q=n,\, p>0} \upsilon_q \star \gamma_p $$
are coboundaries, where
$$  \upsilon_q \star \gamma_p (a,b,c)=\upsilon_q ( \gamma_p(a,b),c) - \upsilon_q(a,\gamma_p(b,c)). $$
In particular, for $\upsilon_i=\phi_i(\gamma) \in C^2(\cA,\cA_i)$ we have 
$\omega_n(\upsilon_i)=\phi_i(\omega_n(\gamma))$, hence if $\gamma$ is an unobstructed deformation
of $\cA$ with $[\omega_n(\gamma)]=0 \in HH^3(\cA,\cA)$ we also have that $\upsilon_i=\phi_i(\gamma)$
is a cocycle that extends to $(Z^*(\cA,\cA_i)[[t]],\delta_\gamma)$. Moreover, the 
cocycles $\upsilon_i=\phi_i(\gamma) \in C^2(\cA,\cA_i)$
determined the same class $[\psi_1(\upsilon_1)]=[\psi_2(\upsilon_2)]\in HH^3(\cA,\cA_3)$ 
by the long exact sequence. The case for the contravariant
functoriality $\phi_i^*: HH^n(\cA_i,\cA_i)\to HH^n(\cA,\cA_i)$ is similar: if $\gamma_i$ is an
unobstructed deformation of $\cA_i$ then the $2$-cocycle $\phi_i^*(\gamma_i)$ extends to
$(Z^*(\cA,\cA_i)[[t]],\delta_\gamma)$.  
\endproof

\smallskip

We can then propose as compatibility condition between the deformations of
the algebras $\cA_i$ and of $\cA$ as the condition that the $2$-cocycles in
$Z^2(\cA,\cA_i)$ obtained in this way define the same class,
$[c(\gamma_i)]=[c_i(\gamma)] \in HH^2(\cA,\cA_i)$.

\smallskip

We still need to discuss how an unobstructed deformation theory for the twistor spaces $Z_i=Z(M_i)$
determines an unobstructed deformations of their blowups $\tilde Z_i={\rm Bl}_{\C\P^1_{x_i}}(Z_i)$.
In fact, this issue is already discussed in \cite{DoFr}, although only deformations in the ``commutative
direction" $HH^{1,1}(\bA,\bA)$ are considered there with obstruction map $\Phi: HH^{1,1}(\bA,\bA)
\to HH^{2,1}(\bA,\bA)$. We show here how the argument needs to be modified 
in our setting to account  for the full non-commutative deformation 
theory in $HH^2(\bA,\bA)$ with obstruction map $\Phi: HH^2(\bA,\bA) \to HH^3(\bA,\bA)$. 

\begin{prop}\label{defZtildeZ}
For $i=1,2$, let $(Z_i,L_i)$ be the pairs of the twistor spaces $Z_i=Z(M_i)$ of (anti)self-dual
Riemannian $4$-manifolds $M_i$ and the twistor lines $L_i=\C\P^2_{x_i}$ over chosen points
$x_i\in M_i$. Let $\gamma_i$ be unobstructed noncommutative
deformations of the pairs $(Z_i,L_i)$. These determine compatible unobstructed
noncommutative deformations of $Z_i$ and of the blowups $\tilde Z_i$.
\end{prop} 

\proof
Let $\gamma_i^{1,1}$ be the Hodge component in 
$HH^{1,1}(\bA(Z_i),\bA(Z_i))$. Since we are assuming that $\gamma_i$ involves a nontrivial
deformation in the noncommutative direction, we know $\gamma_i^{1,1}\neq 0$.
The $\gamma_i \in HH^2(\bA(Z_i),\bA(Z_i))$ satisfy $\Phi(\gamma_i)=0\in HH^3(\bA(Z_i),\bA(Z_i))$
hence the $\gamma_i^{1,1} \in HH^{1,1}(\bA(Z_i),\bA(Z_i))$ also satisfy $\Phi(\gamma_i^{1,1})=0 \in HH^{2,1}(\bA(Z_i),\bA(Z_i))$. We can view the noncommutativ deformation as being parameterized by a nontrivial
holomorphic skew multivector field, $\gamma_i \in H^0(Z_i, \Lambda^2 \cT_{Z_i}))$,
with the obstruction vanishing in $H^0(Z_i, \Lambda^3 \cT_{Z_i}))$. As in \cite{DoFr}, we denote by
$\cT_{Z_i,L_i}$ the sheaf of holomorphic vector fields on $Z_i$ that are tangent to $L_i$ along $L_i$.
These are related to $\cT_{Z_i}$ by the short exact 
sequence of sheaves $0 \to \cT_{Z_i,L_i} \to \cT_{Z_i} \to \nu_i \to 0$, with $\nu_i$ 
the normal bundle of $L_i$ in $Z_i$.
Holomorphic vector fields on $Z_i$ that preserve $L_i$ extend to holomorphic vector fields on the
blowup $\tilde Z_i ={\rm Bl}_{L_i}(Z_i)$, hence deformations of the pair $(Z_i,L_i)$ classified
by elements in $H^0(Z_i, \Lambda^2 \cT_{Z_i, L_i}))$ with obstructions in $H^0(Z_i, \Lambda^3 \cT_{Z_i, L_i}))$
determine corresponding deformations of $\tilde Z_i ={\rm Bl}_{L_i}(Z_i)$. 
Since $Z_i$ is a $3$-dimensional complex manifold, sections in $H^0(Z_i, \Lambda^3 \cT_{Z_i}))$
are spanned as $\cA(Z_i)$-module by $\partial_{z_0}\wedge \partial_{z_1}\wedge \partial_{z_2}$, for
local coordinates $(z_0,z_1,z_2)$.
Since $L_i$ is a line, this means that along $L_i$ the vector fields in $\cT_{Z_i,L_i}$ 
are generated as $\cA(Z_i)$-module by $\partial_z$ with $z$ a local coordinate on 
the line $L_i$, hence the exterior powers vanish along $L_i$, which means that sections
of $\Lambda^3 \cT_{Z_i,L_i}$ are locally of the form 
$f(z_0,z_1,z_2) \, \partial_{z_0}\wedge \partial_{z_1}\wedge \partial_{z_2}$ where
$f$ is in the ideal of functions vanishing along the line $L_i$. Similarly, if $\partial_{z_i}\wedge \partial_{z_j}$
with $i< j$ is a local basis for sections of $\Lambda^2\cT_{Z_i}$, we can see sections of 
$\Lambda^2\cT_{Z_i,L_i}$ as satisfying a vanishing condition along $L_i$. The obstructions
$\omega_n(\gamma_i)$ of a section of $\Lambda^2\cT_{Z_i,L_i}$ are given by
$$ \omega_n(\gamma_i)=\sum_{\ell + k=n,\, \ell>0} \gamma_{i,\ell}\star \gamma_{i,k}, $$
for a collection of Hochschild $2$-cochains $\{ \gamma_{i,k} \}_{k\in \N}$ for the pair $(Z_i, L_i)$.
These determine Hochschild $3$-cocycles of the deformation theory of $(Z_i,L_i)$ which
we can identify with sections in $H^0(Z_i, \Lambda^3 \cT_{Z_i, L_i}))$.
Thus, we obtain that unobstructed noncommutative deformations of the pairs $(Z_i, L_i)$ determine 
compatible unobstructed noncommutative deformations of $Z_i$ and of the blowup $\tilde Z_i$. 
\endproof

\smallskip

In this section we have focused primarily on the Riemannian case, in order to
compare our deformation and gluing procedure for noncommutative twistor spaces,
based on the Gertenhaber--Shack complex, with the deformation and gluing theory
of classical twistor spaces of Donaldson--Friedman, which is formulated in the
Riemannian context. It is important to stress, though, that the Gertenhaber--Shack 
approach to deformations and the associated obstruction theory does not require 
the Riemannian assumption and can be applied very generally to noncommutative
twistor spaces, either Riemannian ot Lorentzian, described in terms of deformation
quatizations. Other forms of noncommutative deformations, such as those based
on the Connes--Landi $\theta$-deformations, however, have an underlying Riemannian
assumption, since they are based on the spectral triples formalism, which at present
is not fully developed in the Lorentzian case. On the other hand, in the case of the
original quantization of twistor spaces of \cite{Pen68} the Gertenhaber--Shack  formalism 
described in this section applies in both Riemannian and Lorentzian setting, and provides
a general setting for the gluing problem described in Section~D of \cite{Penrose2020}.
We discuss these specific cases more in detail in the next section.

\medskip
\section{Gluing Quantized Twistor Spaces}\label{GlueCasesSec}

The deformation and gluing procedure described above is very general in the sense
that it applies in any setting where a quantization of twistor spaces is constructed using
a deformation quantization procedure. The specific quantizations of twistor spaces
that we discussed in the Section~\ref{NCtwistSec}, however, have additional structure such
as the geometric quantization, the $\theta$-deformation, the deformation quantization
of the Hopf fibration, and the almost commutative geometry. Thus, it is better for each
of these cases to analyze how a gluing procedure works that accounts for these
additional structures. 

 \smallskip
 \subsection{Gluing of geometric quantizations}
 
 We start with our main object of interest, which is the geometric quantization of twistor spaces
 constructed by one of us in \cite{Pen68}. We have shown in \S \ref{PenroseQSec} that
 these quantized twistor spaces can be seen as deformation quantizations, through the
 Fedosov relation \cite{Fed} between geometric and deformation quantization.
 We can then apply the construction we presented in \S \ref{DefSec}.
 
 \smallskip
 
 We can proceed as described in the previous section to construct a noncommutative
 twistor space for the connected sum $M=M_1\# M_2$, given the quantizations of
 the twistor spaces $Z(M_i)$. If these quantizations are obtained using the geometric
 quantization method of \cite{Pen68}, then we want to check that, if a classical
 unobstructed deformation $Z_t$ exists of the singular gluing $\tilde Z$ of the blowups $\tilde Z_i$
 of the twistor spaces $Z(M_i)$, then the gluing of the quantized twistor spaces can be
 performed in a way that gives rise of a geometric quantization of the deformation $Z_t$. 
 
 \smallskip
 
\begin{prop}
Let $M_i$ be two (anti)self-dual Riemannian manifolds with $Z_i=Z(M_i)$ their twistor spaces.
Under the connected sum $M=M_1\# M_2$ operation, a gluing of the geometric quantizations 
of the $Z_i$'s is determined by the geometric quantization of a Gompf symplectic sum of 
of the $X_i=\cS^+(M_i)_0$ that fiber over $Z_i$ with $\C^*$ fibers.
\end{prop} 
 
\proof 
As in \S \ref{PenroseQSec}, we consider the symplectic form $\omega_i =\sum_\alpha dZ_i^\alpha\wedge 
d \bar Z_{i,\alpha}$ on $X_i:=\cS^+(M_i)_0$, and we consider $\tilde X_i =\tilde \cS^+(M_i)_0$, the pullback 
of the $\C^*$-bundle $\cS^+(M_i)_0$ along the projection map $\tilde Z_i \to Z_i$ from the blowup
$\tilde Z_i={\rm Bl}_{L_{x_i}}(Z_i)$ of a twistor line $L_{x_i}$ in $Z_i$. The singular space $\tilde Z$
obtained by the gluing of the complex manifolds $\tilde Z_i$ along their exceptional divisors
$\tilde Z=\tilde Z_i \cup_{E_1\simeq E_2} \tilde Z_2$ corresponds to a gluing $\tilde X = \tilde X_1 \cup_{V_1\simeq V_2} X_2$, with $V_i$ the real codimension two symplectic submanifold of $\tilde X_i$ given by the 
preimage of the exceptional divisor $E_i$, which is a singular symplectic variety with a normal crossings
singularity. The singular space $\tilde Z$ satisfies the $d$-semistable condition,
namely the normal bundles $\nu_i$ of $E_i$ inside $\tilde Z_i$ are such that $\nu_1\otimes \nu_2$ is
the trivial line bundle. Thus, the Gompf symplectic sum construction of \cite{Gompf} applies to the
pairs $(\tilde X_i, V_i, \tilde\omega_i)$ and gives a one-parameter deformation family, in the form of
a nearly regular symplectic fibration $(\cX,\omega, \pi: \cX \to \C)$ with $\pi^{-1}(0)=\tilde X$.
For $t\neq 0$, the restriction $\omega_t$ of $\omega$ to $X_t=\pi^{-1}(t)$ is non-degenerate,
and is a smoothing of $\tilde X$.  Thus, we can regard the geometric quantization of $X_t$ 
as the quantized twistor space resulting from the gluing of the quantized twistor spaces of the manifolds $M_i$.
If the classical deformation theory of $\tilde Z$ is unobstructed, so that we have a smooth deformation 
$Z_t$ of $\tilde Z$, then this construction can be done
compatibly with the $\C^*$-fibrations $X_i \to Z_i$ so that $X_t=\cS^+(M_t)_0$ provides such a 
deformation, where $M_t$ is the connected sum $4$-manifold $M_1\# M_2$ endowed
with an (anti)self-dual metric $g_t$ for which $Z_t=Z(M_t)$ is the twistor space. 
\endproof

  \smallskip
 \subsection{Gluing of deformations of the Hopf fibration}
 
 In the case of the deformation quantization of the Hopf fibration
 and the associated quantization of twistor spaces discussed in \S \ref{defqNCtwistSec}
 the question is whether the gluing procedure described in \S \ref{DefSec} maintains
 the compatibility with the Hopf fibration. Since in the Riemannian setting the Hopf fibration $Z(M)\to M$ with
 fibers the twistor lines assumes the existence of an (anti)self-dual structure on $M$,
 we can work under the hypothesis that the underlying commutative deformation
 theory of the $Z(M_i)$ is unobstructed and there is a resulting twistor space
$Z(M)$, where $M=M_1\# M_2$ has an (anti)self-dual structure, obtained as 
classical deformation of the singular $\tilde Z$ as in \cite{DoFr}. 

\smallskip

Under this assumption, we can identify the result of the noncommutative deformation of 
$\tilde Z$ of \S \ref{DefSec} with a noncommutative deformation of $Z(M)$. We need to
check that, if the noncommutative deformations of the $Z(M_i)$ are chosen to be deformations
as in \S \ref{defqNCtwistSec}, obtained via a noncommutative deformation of the Hopf
fibration $S^1\hookrightarrow S^3 \to \C\P^1$, then the resulting noncommutative
deformation of $Z(M)$ is also of this form. We can view this as the noncommutative
analog of the argument of \cite{DoFr} showing that the classical deformation of the
singular space $\tilde Z$ is indeed the twistor space of $M=M_1\# M_2$, hence
in particular it has an associated Hopf fibration. Indeed the result of \cite{DoFr}
for the classical deformation will directly imply the compatibility of the noncommutative
deformations.

\smallskip

\begin{prop}\label{defHopfglue}
Let $Z_{i,\hbar}$ be noncommutative deformations of the twistor spaces $Z_i=Z(M_i)$
with compatible noncommutative deformations $S_{i,\hbar}$ of $S_i=S(M_i)$, 
obtained as in Proposition~\ref{propNCtwistor} that fit in the Hopf fibrations diagram
\eqref{qHopfSZ}. Let $Z_t$ be a classical deformation of the singular 
space $Z_0=\tilde Z$ obtained by gluing the blowups of $Z_i$ at a twistor line along
the exceptional divisors. Then the deformations $Z_{i,\hbar}$ and $S_{i,\hbar}$ and $Z_t$
determine compatible noncommutative deformations $\tilde Z_\hbar$, $\tilde S_\hbar$
and $Z_{t,\hbar}$ and $S_{t,\hbar}$ that satisfy the same compatibility with the
Hopf fibration as in \eqref{qHopfSZ}.
\end{prop}

\proof
As in \cite{DoFr}, notice that the set of $\C\P^1$ lines in the blowup $\tilde Z_i={\rm Bl}_{\C\P^1_{x_i}}(Z_i)$
 is parameterized by $M_1\smallsetminus \{ x_i \} \cup \P(T_{x_i}(M_i))$, that is, the real blowup
 $\tilde M_i ={\rm Bl}_{x_i}(M_i)$, with $\P(T_{x_i}(M_i))\simeq \R\P^3$. The set of  $\C\P^1$ lines in
 the singular space is similarly parameterized by the gluing of these real blowups along the exceptional
 divisors $P_i:=\P(T_{x_i}(M_i))\simeq \R\P^3$, which we denote by $\tilde M=\tilde M_i \sqcup_{P_1\simeq P_2} \tilde M_2$. 
 Thus, we can construct a space $\tilde S$ obtained from the singular space $\tilde Z$ by the Hopf fibration
 diagram
 \begin{equation}\label{tildeSZHopf}
 \xymatrix{
 S^1 \rto^{=} \dto & S^1 \dto &  \\ 
S^3 \rto \dto & \tilde S \rto \dto & \tilde M \dto^{=} \\ \C\P^1 \rto & \tilde Z \rto & \tilde M
 }
 \end{equation}
 where the map $\tilde Z\to \tilde M$ has fiber over $x\in \tilde M$ the $\C\P^1$ line in $\tilde Z$
 that the point $x$ parameterizes, and the space $\tilde S$ is obtained by building over each $\C\P^1$ 
 line in $\tilde Z$ a $3$-sphere $S^3$ via the Hopf fibration. 
 
This allows us to apply the construction of the compatible noncommutative deformations of 
Proposition~\ref{propNCtwistor} to the pair $\tilde S$, $\tilde Z$, compatibly with the
noncommutative deformation of the Hopf fibration, which we represent as the
diagram of noncommutative spaces
\begin{equation}\label{qtildeSZHopf}
 \xymatrix{
 S^1 \rto^{=} \dto & S^1 \dto &  \\ 
S^3_\hbar \rto \dto & \tilde S_\hbar \rto \dto & \tilde M \dto^{=} \\ \C\P^1_\hbar \rto & \tilde Z_\hbar \rto & \tilde M
 }
 \end{equation}

\smallskip

We then consider an unobstructed one-parameter deformation $Z_t$ of the singular space $Z_0=\tilde Z$ to
the twistor space $Z(M)$ of the connected sum manifold $M=M_1 \# M_2$. We
denote by $M_t$ the (anti)self-dual structure on $M$ that is the smoothing of $\tilde M$ with
local form $xy=t$ near the normal crossings singular locus of $\tilde M$. Then, as shown in
\cite{DoFr}, the set of lines of $Z_t$ is parameterized by the points of $M_t$; all lines have 
the correct normal bundle $\cO(1)\oplus \cO(1)$ and a fixed-point-free antiholomorphic involution
leaving the lines invariant, hence they satisfy the characterization of twistor spaces
and can be identified with $Z_t=Z(M_t)$. Thus, we also have an associated $S_t=S(M_t)$
that fits in the Hopf fibrations diagram \eqref{HopfSZ}.
We can then apply the same construction of 
Proposition~\ref{propNCtwistor} on all of the pairs $(S_i,Z_i)$, $(\tilde S_i, \tilde Z_i)$,
$(\tilde S, \tilde Z)$, $(S_t, Z_t)$ and obtain corresponding noncommutative deformations
obtained by deforming the Hopf fibration. The compatibility between all of these
noncommutative deformations comes from the compatibilities of the underlying
commutative spaces parameterizing lines in $Z_i$, $\tilde Z_i$, $\tilde Z$ and $Z_t$.
\endproof  
 
  \smallskip
 \subsection{Gluing of $\theta$-deformations}
 
 In the case of the $\theta$-deformations (as well as the case of the fuzzy twistor spaces
 that we discuss below in \S \ref{fuzzyglueSec}) the gluing procedure can be handled in 
 a different way that does not require relying on the noncommutative
 deformation theory of \S \ref{DefSec}.
 
 \smallskip
 
 \begin{prop}\label{thetadefglue}
 Suppose given unobstructed classical one-parameter deformation $Z_t$ of the singular space 
 $\tilde Z$ to the twistor space $Z(M)$ of the connected sum $M=M_1\# M_2$.  This determines
 an associated family of $\theta$-deformations $S(M_t)$ compatible with a $\theta$-deformation
 $\tilde S_\theta$ of a fibration $\tilde S$ over the singular $\tilde Z$ and $\theta$-deformations
 $S(M_i)_\theta$ obtained as in \S \ref{thetadefSec}.
 \end{prop}

 \proof
 In the case of the $\theta$-deformations of \S \ref{thetadefSec} it is only the spaces $S(M)_\theta=\bS(\cS^+(M))_\theta$ that is deformed to a noncommutative space, while the twistor space $Z(M)$ itself remains
 commutative. In this case, the gluing and deformation theory of $Z(M_i)$ and $Z(M_2)$ remains
 the same as in the setting of \cite{DoFr}, with the twistor space $Z(M_1\# M_2)$ obtained from
 a commutative deformation of the singular space $\tilde Z$ built from unobstructed commutative
 deformations of the $Z(M_i)$. Thus, in order to obtain compatible noncommutative deformations
 of the $S(M_i)_\theta$ that glue to a noncommutative deformation of $S(M)_\theta$, we need to
 show how to associate to the choice of an unobstructed deformation of $Z(M_i)$ and the
 noncommutative $\theta$-deformations $S(M_i)_\theta$ a 
 resulting deformation of the singular space $\tilde Z$ with an associated
 $\theta$-deformation $S(\tilde Z)$ such that the deformation of $\tilde Z$ to $Z(M)$ yields
 the desired $\theta$-deformation $S(M)_\theta$.
 
 \smallskip
 
 As in Proposition~\ref{defHopfglue}, we parameterize lines in $\tilde Z_i$ by 
 the real blowp $\tilde M_i$ and lines in $\tilde Z$ by the resulting gluing $\tilde M$,
 and the construct compatible fibrations $\tilde S_i$ and $\tilde S$ that fit the
 Hopf fibration diagram \eqref{tildeSZHopf}. 
  This leads to an associated construction
 of a $\theta$-deformation $\tilde S_\theta$ that fits the noncommutative Hopf fibration diagram
 \begin{equation}\label{tildeSZHopftheta}
 \xymatrix{
 S^1 \rto^{=} \dto & S^1 \dto &  \\ 
S^3_\theta \rto \dto & \tilde S_\theta \rto \dto & \tilde M \dto^{=} \\ \C\P^1 \rto & \tilde Z \rto & \tilde M
 }
 \end{equation}
 with $\tilde Z$ corresponding to a $U(1)$-invariant subalgebra of the algebra $\cA(\tilde S_\theta)$
 of the $\theta$-deformation, as in the cases of the $\theta$-deformations $S(M_i)_\theta$ 
 obtained as in \S \ref{thetadefSec}. The compatibility between the $\theta$-deformations 
 $\tilde S_\theta$ and the $S(M_i)_\theta$, for $i=1,2$ is provided by the fact that the parameterizing
 families of lines of $\tilde Z$ in the complement of the glued exceptional divisors agree with those
of $Z_i$, namely with $M_i\smallsetminus \{ x_i \}$ so that the $\theta$-deformations of the
$3$-spheres over each of these $\C\P^1$-lines agree for $S(M_i)$ and $\tilde S$. 

\smallskip

We then proceed again as in Proposition~\ref{defHopfglue}, by considering the
$Z_t$ and compatible $S_t$ that fit the Hopf fibration diagrams \eqref{HopfSZ}.
The characterization of \cite{DoFr} of $Z_t$ as twistor spaces of the (anti)self-dual structure
$M_t$ on the connected sum spacetime manifold then identifies the $S_t$ constructed in
this way with $S(M_t)$. 

This means that we can then compatibly build a family of $\theta$-deformations $S(M_t)_\theta$
that fit into the Hopf fibrations diagram \eqref{thetaHopfSZ}. 
\endproof

\smallskip

The gluing of the $\theta$-deformations of Proposition~\ref{otherThetaDef} is more
interesting, since in this case the twistor spaces $Z(M_i)$ themselves are deformed
to noncommutative spaces $Z(M_i)_\theta$ via a $\theta$-deformation of the
respective subspaces $\P N_i \subset Z(M_i)$ along the Hopf spheres $S^3_\theta$.

\smallskip

\begin{prop}\label{glueotherTheta}
The gluing $\tilde Z$ of the blowups $\tilde Z(M_i)={\rm Bl}_{\C\P^1_{x_i}}(Z(M_i))$ of the 
twistor spaces $Z(M_i)$ along the exceptional divisors $E_i$ admits a $\theta$-deformation
$\tilde Z_\theta$ compatible with the $\theta$-deformations $Z(M_i)_\theta$ of Proposition~\ref{otherThetaDef}.
\end{prop}

\proof Since in the $\theta$-deformations of Proposition~\ref{otherThetaDef} only the subspace
$\P N$ of the twistor space is deformed to a noncommutative space, we can assume that
the points $x_i\in M_i$ are chosen so that the fibers $\C\P^1_{x_i}$ are contained in the respective
$\P N_i \subset Z(M_i)$. The Hopf spheres that are $\theta$-deformed to obtain the
noncommutative $Z(M_i)_\theta$ are the intersections $S^3_P=P\cap \P N_i$ with a family of planes
passing through a chosen point in the positive norm part $\P T^+$ of the twistor space. 
The planes $P$ cut out Hopf circles $S^1_{x_i,P}$ in the fiber $\C\P^1_{x_i}$. Since the
noncommutative deformation involves the individual Hopf spheres $S^3_P$, we can restrict
our attention to a single sphere. Thus, instead of working with the gluing $\tilde Z$ of the
blowups $\tilde Z(M_i)={\rm Bl}_{\C\P^1_{x_i}}(Z(M_i))$ along their exceptional divisors
we can consider Hopf spheres $S^3_{P_i}\subset \P N_i$ and their 
real blowups along the Hopf circles $S^1_{x_i,P_i}$. The resulting singular space $S$
is a gluing of two $3$-spheres $S^3_{P_i}$ along a torus $T^2$, which can be identified with 
the boundary of a tubular neighborhood of one of the Hopf circles $S^1_{x_i,P_i}$. 
In Hopf coordinates, we can identify such a tubular neighborhood with values $0\leq \eta \leq \epsilon$
for some $\epsilon>0$ and $(\xi_1,\xi_2)\in T^2$. Thus, the resulting space can be regarded
as the gluing of two copies of $S^3$ along one of the tori $T^2$ that are $\theta$-deformed to
noncommutative tori in the deformation to $S^3_\theta$. While the gluing $S$ of the two $3$-spheres
along the torus is not a smooth manifold, but has a normal crossing singularity along the torus,
it is still possible to define a $\theta$-deformation $S_\theta$, since the deformation happens
along the individual tori that are deformed to noncommutative tori. These are either
the torus along which the gluing is performed or the other Hopf tori in each of the to $S^3$.
The resulting $\theta$ deformation is obtained by considering the algebra of
functions that are smooth on each of the two $S^3$ and that have matching values on
the torus where the gluing is performed. The noncommutative $\theta$-deformed product
on this algebra of functions is then defined as in \eqref{thetadefprod}, which has the
effect of deforming all the individual Hopf tori in $S$ to noncommutative tori. In order to
view the $\theta$-deformation $S_\theta$ as a spectral triple, one can take as 
Hilbert space and Dirac operator the direct sum of the respective ones on the two 
copies of $S^3$. This is analogous to the spectral triple construction used in the
gluing of copies of smooth manifolds into fractal configurations, see \cite{CCL},
\cite{FKM}. 
\endproof

\smallskip

In this case, because of the very explicit nature of the noncommutative
deformation in terms of Hopf tori, we have not used the description of
deformations in terms of Hochschild cohomology. Notice, however, that
a description of the deformation and obstruction theory for the noncommutative
$3$-spheres $S^3_\theta$ has been discussed, in a more general setting
of noncommutative deformations of $3$-spheres, by Connes and Dubois-Violette
in \cite{CoDV}.

 \smallskip
 \subsection{Gluing of fuzzy twistor spaces}\label{fuzzyglueSec}
 
 The fuzzy twistor spaces introduced in \S \ref{fuzzySec} provide an example
 of quantization of twistor spaces to which the general deformation theory
 approach described in \S \ref{DefSec} does not directly apply, due to a
 rigidity property. 

 \begin{prop}\label{fuzzyrigid}
 The fuzzy twistor spaces are rigid, in the sense that they do not admit
 any deformations that maintain the underlying spacetime manifold $M$
 commutative. 
  \end{prop}
 
 \proof
 It is shown in \S 16 of \cite{GeSch} that the Hochschild cohomology and the
set of equivalence classes of deformations are a Morita invariant. Namely, if
two unital associative algebras $\cA$ and $\cB$ are Morita equivalent, then
there is an isomorphism $HH^*(\cA,\cA) \to HH^*(\cB,\cB)$ that preserves the cup product
and the graded Lie bracket. A bijection between the set of equivalence classes of
deformations is then obtained in the following way. There is a finitely generated
projective right $\cA$-module $\cE$ such that $\cB={\rm End}_\cA(\cE)$, hence
we can identify $\cB=e M_N(\cA) e$,
for some idempotent $e \in M_N(\cA)$ with $M_N(\cA) e M_N(\cA) =M_N(\cA)$.
by regarding $\cE$ as a summand of a free module of rank $N$. Given a deformation
$\cA_t$ of $\cA$, there is a corresponding deformation $M_N(\cA_t)$ of $M_N(\cA)$ and
an idempotent $e_t$ in $M_N(\cA_t)$ with constant term equal to $e$. Then 
$e_t M_N(\cA_t) e_t$ determines a deformation of $\cB$ that is Morita equivalent to $\cA_t$. 

\smallskip

For a Fr\'echet algebra of smooth functions on a compact smooth manifold, $\cA=\cC^\infty(M)$,
the behavior of the Hochschild homology was analyzed in \cite{Co}. As discussed in \cite{Nada},
the continuous and smooth deformation theories of $\cC^\infty(M)$ are governed by the same
cohomology $HH^*_{cont}(\cC^\infty(M),\cC^\infty(M))\simeq HH^*_{smooth}(\cC^\infty(M),\cC^\infty(M))
\simeq H^0(M,\Lambda^*\cT_M)$, where $H^0(M,\Lambda^*\cT_M)$ denotes the space of global
sections of the exterior algebra of the tangent bundle of $M$, that is, the skew multivector fields on $M$. 
The infinitesimal deformations up to equivalence can then be seen as the elements of the second 
Hochschild cohomology, that is, the sections in $H^0(M,\Lambda^2\cT_M)$, while the obstructions
live in this third cohomology, identified with $H^0(M,\Lambda^3 \cT_M)$. These correspond only
to deformations of the spacetime manifold $M$ in the ``noncommutative direction". Thus,
all these deformations violate the requirement that spacetime itself remains commutative. 
\endproof 

In this case, however, a gluing of fuzzy twistor spaces that corresponds to the connected sum
of the underlying spacetime manifolds can be performed without passing through the
deformation and obstruction theory discussed in \S \ref{DefSec}. 

\begin{prop}\label{gluingfuzzy}
The orientation reversing isometry $\gamma: T_{x_1} M_1\to T_{x_2} M_2$ at the
points $x_i \in M_i$ where the connected sum $M= M_1\# M_2$ is performed determines
a gluing of the fuzzy twistor spaces of $M_i$ to a fuzzy twistor space of $M$.
\end{prop}

\proof
Consider the manifolds with boundary $M'_i$, obtained by removing a small ball $U_i$ near the points $x_i$,
with boundary $\partial M'_i\simeq S^3$. We can consider cylindrical ends $S^3\times [0,\epsilon)$ 
attached to $M'_i$ and a metric (without self-duality property) interpolating smoothly between the metric
of $M'_i$ to a cylindrical metric on a smaller interval, built using the metric on the tangent space
at $x_i$. The orientation reversing isometry $\gamma: T_{x_1} M_1\to T_{x_2} M_2$ determines a 
gluing map that identifies these cylindrical ends. Assuming that over a slightly larger ball $U'_i \subset M_i$
containing $x_i$ the almost commutative geometry of the fuzzy space is a product 
$\cA(U'_i)\otimes \cA(S^2_N)$, we can use the same $SU(2)$-valued gluing map, seen as an 
automorphism of $\cA(S^2_N)$ through the $(N+1)$-dimensional representation of $SU(2)$, to
glue together the noncommutative spaces $\cA(S^2_N)$ over the cylindrical ends. The resulting
noncommutative space is still an almost commutative geometry, in the general form of \cite{Cac2}
where the underlying commutative geometry is by construction $M$, with noncommutative fiber
$S^2_N$, hence it provides a model for a fuzzy twistor space for $M$, regardless of the existence 
of an (anti)self-dual metric on the connected sum.
\endproof

\medskip
\subsection*{Acknowledgment} The first author is partially supported by
NSF grant DMS-1707882, and by NSERC Discovery Grant RGPIN-2018-04937 
and Accelerator Supplement grant RGPAS-2018-522593.

\end{document}